\documentclass[final,12pt]{elsarticle}

\usepackage{amssymb}
\usepackage{amsmath}

\usepackage{array}
\usepackage{natbib}
\usepackage{geometry}
\usepackage{graphicx}
\usepackage{xcolor}
\usepackage{listings}
\usepackage{subcaption}
\usepackage{hyperref}
\usepackage{cleveref}
\hypersetup{
    hypertexnames = false,
    colorlinks = true,
    urlcolor   = blue,
    citecolor  = black,
}
\newcommand{\cvec}[1] {\ensuremath{\boldsymbol{ \mathcal{ #1 }}}}

\journal{Computer Physics Communications}

\begin{document}

\begin{frontmatter}



\title{DeboRheo v2.0: Improved Robustness for Geometrical VOF Simulations of Viscoelastic Two-Phase Flows}


\author[FMI]{Daulet Izbassarov}
\affiliation[FMI]{organization={Finnish Meteorological Institute},
            addressline={Erik Palmenin aukio 1}, 
            city={Helsinki},
            postcode={00560}, 
            country={Finland}}

\author[TUD]{Matthias Niethammer}
\affiliation[TUD]{organization={Technical University of Darmstadt},
            addressline={Mathematical Modeling and Analysis}, 
            city={Darmstadt},
            postcode={64287}, 
            country={Germany}}

\author[FMI]{Rahul Subburaj}

\author[TUD]{Dieter Bothe}

\author[UL]{Francesco Romanò}
\author[UL]{Renjie Hao}
\affiliation[UL]{organization={Université de Lille, CNRS, ONERA, Arts et Métiers Institute of Technology, Centrale Lille},
            addressline={UMR 9014 -LMFL - Laboratoire de Mécanique des Fluides de Lille - Kampé de Fériet}, 
            city={Lille},
            postcode={F-59000}, 
            country={France}}
            
\author[KU]{Metin Muradoglu}
\affiliation[KU]{organization={Koc University},
            addressline={Department of Mechanical Engineering}, 
            city={Istanbul},
            postcode={34450}, 
            country={Turkey}}

\author[TUD]{Tomislav Maric}

\begin{abstract}


We present DeboRheo v2.0, an unstructured, geometrically unsplit
Volume-of-Fluid framework implemented in OpenFOAM for viscoelastic two-phase
flows. Version 2.0 improves numerical robustness when large viscosity contrasts
coincide with strongly deformed interfaces and thin viscoelastic lamellae or
filaments. The revised interfacial viscosity treatment retains a sharp
geometric interface while regularizing the viscous contribution to the
momentum equation. The solver combines mass--momentum-consistent phase
advection, optional SIMPLEC pressure--velocity coupling, and a modular
rheological framework with conformation-tensor stabilization. We assess the
method using two-dimensional, axisymmetric, and three-dimensional benchmarks
governed by the Oldroyd-B and exponential Phan--Thien--Tanner models. The
results agree well with published numerical and experimental data and
demonstrate improved robustness for demanding viscoelastic free-surface
configurations on unstructured meshes.

\noindent{\bf PROGRAM SUMMARY}
\begin{small}
\noindent
{\em Program Title:} DeboRheo V2.0                                       \\
{\em CPC Library link to program files:} (to be added by Technical Editor) \\
{\em Developer's repository link:} \url{https://gitlab.com/deborheo/deborheorelease/-/tree/v2.0.5} \\
{\em Code Ocean capsule:} (to be added by Technical Editor)\\
{\em Licensing provisions:} GPLv3 \\
{\em Programming language:} C++\\
{\em Nature of problem:}
Direct numerical simulation (DNS) of viscoelastic two-phase flows is challenged
by sharp interfacial discontinuities in material properties and rheological
response. These difficulties are amplified at high Weissenberg numbers and
large viscosity contrasts, particularly when evolving free surfaces form thin
viscoelastic lamellae and filaments that must remain resolved over only a few
computational cells.

\sloppy
{\em Solution method:}
An unstructured, geometrically unsplit Volume-of-Fluid method is presented for
two-phase flows with a viscoelastic liquid phase governed by general rate-type
constitutive models. The approach extends a previously developed VOF solver by
introducing a regularized treatment of dynamic viscosity in the interfacial
region while retaining a sharp geometrically reconstructed interface. This
treatment improves robustness in flows involving large viscosity contrasts and
thin viscoelastic lamellae and filaments. A log-conformation formulation is
employed to mitigate the High Weissenberg Number Problem (HWNP), and optional
SIMPLEC pressure--velocity coupling is available through the OpenFOAM PIMPLE
controls.

\end{small}
\end{abstract}

\begin{graphicalabstract}
\begin{figure}[ht!]
\centering
\input{ShapeBo5p7}
\input{ShapeBo7p4}
\end{figure}
\end{graphicalabstract}


\begin{highlights}
\item DeboRheo v2.0 is developed for robust viscoelastic two-phase flow simulation.
\item Face viscosity is regularized while the geometric VOF interface remains sharp.
\item The treatment improves robustness for thin viscoelastic lamellae and filaments.
\item SIMPLEC is implemented for robust pressure--velocity coupling.
\item The framework is validated using two- and three-dimensional benchmark flows.
\end{highlights}


\end{frontmatter}



\section{Introduction}
\label{sec:intro}


Non-Newtonian fluids play a crucial role in various industrial and natural processes. These fluids are used in the drilling mud for oil and gas \cite{Rafieefar2021}, food processing industries \cite{Rao2010}, and pharmaceutical industries. Most of all, the greatest relevance is found in everyday human items, such as paints, cosmetics, and foods such as ketchup, mayonnaise, etc. \cite{Rao2010}. In nature, non-Newtonian behavior is observed in blood \cite{errill1969blood}, mucus that covers the lung airways \cite{grotberg2011respiratory,Romanojfm2021,Erken2023,Fazla2024}, synovial fluid responsible for joint lubrication \cite{levick1998synovial}, lava, and even egg albumin \cite{mine1995egg}. Due to the substantial differences between complex and Newtonian fluids, there is a compelling need to model and investigate non-Newtonian flows. Some exotic behaviors of non-Newtonian fluids include the Weissenberg effect, memory effects, viscoelasticity, and plasticity. Most of the processes and natural occurrences mentioned above involve fluid interfaces where one or both phases are viscoelastic, just like in the case of the mucus lining in the lung airways~\cite{grotberg2011respiratory}. In most cases, these interfaces are subjected to shear stress, leading to stretching or rupture into droplets~\cite{Subburaj2026}. Thus, accurate treatment of the fluid interfaces is crucial for modeling these processes.   

The simulation of viscoelastic fluids is a fundamentally challenging task due to their highly non-linear stress–strain relationship, particularly at high Weissenberg numbers~\cite{Renardy2000}. In this regime, the large separation between viscous and elastic time scales leads to the well-known High Weissenberg Number Problem (HWNP), which manifests as numerical instabilities in simulations of viscoelastic flows. The HWNP is characterized by exponential stress growth, which leads to severe numerical instabilities in standard formulations. Conventional viscoelastic models such as Oldroyd-B~\cite{Oldroyd1950} can exhibit unbounded stress growth in regions of strong extension or near stagnation points, often leading to numerical instabilities or divergence of the solution. In contrast, finitely extensible nonlinear elastic models (FENE-P~\cite{Bird1980}, FENE-CR~\cite{Chilcott1988}) are designed to prevent unbounded stress growth by imposing a finite extensibility limit. However, they may still exhibit steep but bounded stress growth, which can challenge numerical solvers. Several robust numerical techniques have been developed to address the HWNP~\cite{Fattal2004, Niethammer2018, Alves2021}. While an accurate and stable solution is required even for Newtonian flows, viscoelastic simulations impose additional constraints, notably the preservation of the positive definiteness of the conformation tensor~\cite{Hulsen1990, owens2002computational}. Fattal and Kupferman~\cite{Fattal2004} proposed the log-conformation method (LCM), which reformulates the constitutive equations using a logarithmic transformation of the conformation tensor. This approach significantly enhances numerical stability by ensuring positive definiteness and enabling an accurate capture of sharp viscoelastic stress gradients, even at high Weissenberg numbers. Alternatively, Balci et al.~\cite{Balci2011} proposed a method based on the square root representation of the conformation tensor to achieve the same goal. Comprehensive overviews of stabilization techniques for viscoelastic flows are available in the review publications by Chen et al.~\cite{chen2013comparison} and Alves et al.~\cite{Alves2021}.

Simulating two-phase viscoelastic flows presents significant numerical challenges. Beyond the inherent HWNP common to viscoelastic fluids, these simulations must also resolve steep gradients in shear stresses and abrupt changes in density and viscosity across the fluid interface. These combined factors can severely compromise the stability and accuracy of numerical solutions. While various multiphase methods - including front-tracking~\cite{Izbassarov2015,Naseer2023,Erken2023,Fazla2024}, level-set~\cite{Izbassarov2018}, and phase-field~\cite{Bazesefidpar2022} - have been applied to viscoelastic two-phase flows, the Volume-of-Fluid method is particularly well suited to such problems. Its inherent mass conservation and robustness make it a preferred choice for handling interfacial complexities~\cite{Chinokya2005, Figueiredo2016, LopezHerrara2019}. OpenFOAM, widely adopted open source CFD framework, provides a range of tools to simulate single- and two-phase viscoelastic flows~\cite{Favero2010, Habla2012, Holmes2012, Pimenta2017, Yuan2020, Dritselis2022}. However, most of its available VOF solvers for viscoelastic two-phase systems utilize algebraic VOF methods (such as interFoam), in which volume fractions are artificially smeared over several computational cells. Artificial smearing of volume fractions exceedingly limits its ability to accurately capture the hydrodynamics of thin viscoelastic liquid filaments. Geometric VOF methods have been well documented to offer superior accuracy in interface representation compared to algebraic approaches~\cite{Maric2020}. For example, the widely used Basilisk solver~\cite{Basilisk} employs a geometric VOF technique, representing the interface as a sharp surface element, thus maintaining high accuracy and sharpness in both 2D and 3D structured meshes. Basilisk includes a dedicated solver module for two-phase viscoelastic flow simulations~\cite{LopezHerrara2019}, which has been used in various applications~\cite{esposito2024buoyancy,Hao2025,Hao2026}.

We have recently developed an un-split geometric Volume-of-Fluid approach for simulating viscoelastic two-phase flows based on the one-fluid formulation~\cite{Niethammer2025}. Specifically, we incorporated the piecewise-linear interface calculation with reconstruction of distance functions (plicRDF-isoAdvector), utilizing the geometric VOF libraries provided by twoPhaseFlow~\cite{Scheufler2021}, into the DeboRheo framework~\cite{Niethammer2019,Niethammer2025}. This methodology was validated for a comprehensive set of benchmark cases~\cite{Niethammer2025}. Moreover, our implementation is compatible with a variety of viscoelastic constitutive models and stabilization strategies within the OpenFOAM ecosystem, enhancing its flexibility and applicability to a broad spectrum of two-phase flow scenarios~\cite{Niethammer2019, Niethammer2019b, Meburger2021, Niethammer2025}.


In the present work, we introduce DeboRheo v2.0 and address a numerical limitation encountered when the sharp geometric VOF representation is applied to strongly deformed viscoelastic two-phase flows. In DeboRheo v1.0, the solvent viscosity at an interfacial face was evaluated directly from the geometrically reconstructed face area fraction. Although this treatment is consistent with the sharp geometric representation of the phase indicator, we found it to become numerically sensitive when thin lamellae or filaments approach the local mesh scale, particularly in the presence of strong viscosity contrasts. DeboRheo v2.0 therefore decouples the geometric sharpness of the interface from the discretization of the viscous coefficient: the interface continues to be reconstructed geometrically, whereas the face viscosity entering the momentum equation is obtained from finite-volume interpolation of cell-centered phase information. This introduces a mesh-scale regularization of the viscosity coefficient without smearing the VOF interface itself. In addition to describing this modification, we document the solver architecture and current limitations and assess the methodology against two-dimensional, axisymmetric, and three-dimensional viscoelastic two-phase benchmark problems.

\section{Mathematical model}
\label{sec:model}

\begin{figure}[!htb]
    \centering
    \includegraphics[width=0.6\textwidth]{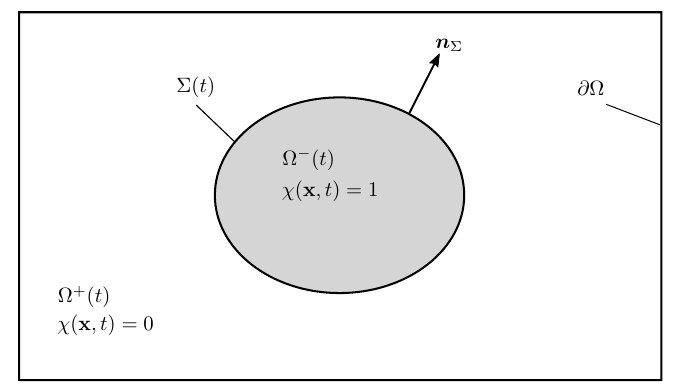}
    \caption{Flow domain $\Omega$, separated by the fluid interface $\Sigma(t)$ into two sub-domains $\Omega^\pm(t)$.}
    \label{fig:model}
\end{figure}

We model incompressible two-phase flows without phase change using the one-field formulation of the Navier-Stokes equations, in which one phase is a viscoelastic fluid. The scalar phase indicator function $\chi(\mathbf{x},t) \in [0,1]$ is used to indicate fluid phases in the flow domain $\Omega$, occupying sub-domains $\Omega^\pm(t) \subseteq \Omega$, as depicted in \cref{fig:model}. It is defined as
\begin{equation}
  \chi(\mathbf{x},t) := 
    \begin{cases}
      1, & \mathbf{x} \in \Omega^-(t), \\ 
      0, & \mathbf{x} \in \Omega^+(t). 
    \end{cases}
  \label{eq:indicator}
\end{equation}
The phase-specific properties of the fluid flow and the physical properties of the individual fluid phases $\phi^\pm(\mathbf{x},t)$ are combined into single fields in the formulation of a field using the phase indicator $\chi$ (\cref{eq:indicator}). 

The one-field formulation of mass density and dynamic viscosity defines
\begin{equation}
    \rho(\mathbf{x},t) = (\rho^-  - \rho^+) \chi(\mathbf{x},t) + \rho^+, \label{eq:rhoindicator} 
\end{equation}
\begin{equation}
    \mu(\mathbf{x},t) = (\mu^- - \mu^+) \chi(\mathbf{x},t) + \mu^+, \label{eq:nuindicator} 
\end{equation} 
\begin{equation}
    \eta(\mathbf{x},t) = (\eta^- - \eta^+) \chi(\mathbf{x},t) + \eta^+, \label{eq:etaindicator}
\end{equation} 
with $\rho^\pm$ as constant mass densities, $\mu^\pm$ as constant dynamic solvent viscosities, and $\eta^\pm$ as constant polymeric viscosities in $\Omega^\pm(t)$.

Similarly to the one-field formulation of mass density and dynamic viscosity, we define the one-field formulation for the polymer stress tensor $\boldsymbol{\tau}$ as 
\begin{equation}
    \boldsymbol{\tau} (\mathbf{x},t) = (\boldsymbol{\tau}^{-}(\mathbf{x},t) - \boldsymbol{\tau}^{+}(\mathbf{x},t)) \chi(\mathbf{x},t)
 + \boldsymbol{\tau}^{+}(\mathbf{x},t).
 \label{eq:onefieldtau}
 \end{equation}
 Unlike phase-specific densities $\rho^\pm$ and dynamic viscosities $\mu^\pm$, the phase-specific polymer stress tensor is generally not constant in space, i.e., $\boldsymbol{\tau}^{\pm}=\boldsymbol{\tau}^{\pm}(\mathbf{x},t)$. We currently limit our two-phase DeboRheo method \citep{Niethammer2025} to a single viscoelastic fluid phase, say 
\begin{equation}
    \boldsymbol{\tau}^{+}(\mathbf{x},t)= \boldsymbol{0},
    \label{eq:taupluszero}
\end{equation}
with no loss of generality concerning which fluid phase has viscoelastic properties as $+$ or $-$ is arbitrarily chosen. With \cref{eq:taupluszero}, \cref{eq:onefieldtau} reduces to  
\begin{equation}
    \boldsymbol{\tau}(\mathbf{x},t) = \boldsymbol{\tau}^{-}(\mathbf{x},t)\chi(\mathbf{x},t).
 \label{eq:finalonefieldtau}
 \end{equation}

The fluid flow in DeboRheo is governed by the one-field formulation of the Navier-Stokes equations for incompressible two-phase flows with one viscoelastic phase, and without phase change, i.e.,
\begin{align}
    \nabla\cdot\mathbf{v} &= \, 0 ,\label{eq:volume-transport} \\
    \partial_t(\rho \mathbf{v})+\nabla\cdot(\rho \mathbf{v} \otimes \mathbf{v})
    - \nabla \cdot \left[ (\mu+\gamma\eta)\nabla\mathbf{v}\right] 
    &=  -\nabla p - (\mathbf{g}\cdot \mathbf{x}) \nabla \rho  \nonumber \\ 
      -\nabla \cdot \left(\gamma\eta\nabla\mathbf{v}\right)
     +
       \nabla \cdot ( \mu\nabla\mathbf{v}^T) 
    & + \mathbf{f}_\Sigma 
     + \nabla \cdot \boldsymbol{\tau},
     \label{eq:momentum-transport}
\end{align}
where $\mathbf{v}$, $p$, $\mathbf{g}$, $\mathbf{x}$, and  $\mathbf{f}_\Sigma$ are the velocity vector, the pressure, the gravitational acceleration vector, the position vector, and the surface tension force, respectively. We add the diffusive term $\nabla\cdot(\gamma\eta\nabla\mathbf{v})$ to both sides of the momentum equation: it is discretized implicitly on the left-hand side and explicitly on the right-hand side, with its magnitude controlled by the scalar parameter $\gamma$. This stabilization technique is known as both-sides diffusion (BSD) \citep{guenette1995new, pimenta2017stabilization, fernandes2017improved}. BSD splits the contribution of the polymeric viscosity \(\eta\), already contained in the polymeric
stress \(\boldsymbol{\tau}\), into explicit and implicit contributions. For example, in the Oldroyd--B model,
\begin{equation}
\boldsymbol{\tau}
+
\lambda
\left(
\frac{\partial \boldsymbol{\tau}}{\partial t}
+
(\mathbf{v}\cdot\nabla)\boldsymbol{\tau}
-
(\nabla\mathbf{v})^{T}\cdot\boldsymbol{\tau}
-
\boldsymbol{\tau}\cdot\nabla\mathbf{v}
\right)
=
\eta
\left(
\nabla\mathbf{v}
+
(\nabla\mathbf{v})^{T}
\right).
\end{equation}
so the polymeric viscous response enters the momentum equation through
\(\nabla\cdot\boldsymbol{\tau}\).

In the segregated solution algorithm, \(\nabla\cdot\boldsymbol{\tau}\) is
treated explicitly. The BSD stabilization adds and subtracts a
Newtonian-like diffusion term,
\[
-\nabla\cdot\!\left[(\mu+\gamma\eta)\nabla\mathbf v\right]
\quad\text{and}\quad
-\nabla\cdot\!\left(\gamma\eta\nabla\mathbf v\right),
\]
as written in the momentum equation. Equivalently, the polymeric stress
contribution is split as
\[
\nabla\cdot\boldsymbol{\tau}
=
\nabla\cdot(\gamma\eta\nabla\mathbf v)
+
\nabla\cdot\!\left(
\boldsymbol{\tau}-\gamma\eta\nabla\mathbf v
\right).
\]
The first part is treated implicitly through the momentum operator, while
the residual remains explicit. Hence the implicit diffusion coefficient is
effectively \(\mu+\gamma\eta\), but the converged equation will be the
original viscoelastic momentum equation.


The two-phase Reynolds transport theorem confirms that the one-field formulation of the velocity 
\begin{equation}
    \mathbf{v} = (\mathbf{v}^- - \mathbf{v}^+) \chi + \mathbf{v}^+
    \label{eq:onefieldv}
\end{equation}
satisfies \cref{eq:volume-transport} if both fluid phases are incompressible, i.e., $\nabla \cdot \mathbf{v}^\pm = 0$, and the relative normal velocity is zero $(\mathbf{v}^- - \mathbf{v}^+) \cdot \mathbf{n}_\Sigma = 0$ at $\Sigma(t)$, i.e., there is no phase change.

Following the addition of forces, the viscoelastic stress $\boldsymbol{\tau}$ is superimposed onto solvent stress, included in \cref{eq:momentum-transport} as Newtonian viscous stress.
In \cref{eq:momentum-transport} we separate the hydrostatic contribution to the pressure into $-(\mathbf{g}\cdot\mathbf{x})\nabla \rho$ as this ensures that the force resulting from the hydrostatic pressure can be balanced with other forces on the fluid interface for stationary fluid interface in the discretization \citep{Liu2024nonorth}. We model the surface tension force using the Continuum Surface Force (CSF) model \citep{Brackbill1992} in a force-balanced formulation \citep{Renardy2001} (cf. \citep{Popinet2018} for a recent review), i.e.,
\begin{equation}
       \mathbf{f}_\Sigma  = \sigma \kappa \mathbf{n}_\Sigma \delta_\Sigma = -\sigma \kappa \nabla \chi.
       \label{eq:fsigma}
\end{equation}
In \cref{eq:fsigma}, 
\begin{equation}
    \kappa = - \nabla_\Sigma \cdot ( \mathbf{n}_\Sigma)
\end{equation}
is twice the mean curvature of $\Sigma(t)$ with $\mathbf{n}_\Sigma$ as the interface normal.

The phase indicator function $\chi$ is advected by the Volume-of-Fluid (VOF) equation, resulting directly from the mass conservation equation and \cref{eq:volume-transport} \citep{Liu2024}, given in integral form as 
\begin{equation}
    \partial_t \alpha_c (t) = - \frac{1}{|\Omega_c|}\int_{\partial \Omega_c} \chi(\mathbf{x},t) \mathbf{v} \cdot \mathbf{n} \, dS ,
    \label{eq:vof}
\end{equation}
for a fixed control volume $\Omega_c$. The integral form in \cref{eq:vof} is conservative by the definition of $\alpha_c$, i.e.,
\begin{equation}
    \label{eq:alphacdef}
    \alpha_c(t)=\frac{1}{|\Omega_c|}\int_{\Omega_c} \chi(\mathbf{x}, t) \, dV = \frac{1}{|\Omega_c|}\int_{\Omega_c^-(t)} 1 \, dV.
\end{equation}
and by the application of the Reynolds transport theorem, which relates the total rate of change of volume (fraction) with the flux through the boundary of the control volume $\partial \Omega_c$. Details on the volume fraction equation and its derivation can be found in \citep{Maric2020,Liu2023}. 


\subsection{Phase-specific viscoelastic stress transport}

The phase-specific nature of the viscoelastic stress, present in DeboRheo only in one fluid phase, requires careful consideration.  Although the framework allows for multimode relaxation times, a single relaxation time is assumed in the present derivation to simplify the analysis.

\begin{table}[ht!]
\centering
\footnotesize
\begin{tabular}{|l|c|c|c|c|c|c|}
\hline
\textbf{Model} & $\zeta$ & $h_0$ & $h_1$ & $g_0$ & $g_1$ & $g_2$ \\
\hline
Maxwell/Oldroyd-B   & $0$ & $-1$ & $1$ & $1$ & $-1$ & $0$ \\
Giesekus & $0$ & $-1$ & $1$ & $1-\alpha$ & $2\alpha-1$ & $-\alpha$ \\
FENE-P              & $0$ & $-1$ & $\frac{1}{1-\mathrm{tr}(\mathbf{C})/b}$ & $1$ & $-h_1$ & $0$ \\
FENE-CR             & $0$ & $-\frac{1}{1-\mathrm{tr}(\mathbf{C})/b}$ & $-h_0$ & $-h_0$ & $h_0$ & $0$ \\
Johnson-Segalman    & $[0,2]$ & $-1$ & $1$ & $1$ & $-1$ & $0$ \\
LPTT                & $[0,2]$ & $-1$ & $1$ & $1 + \frac{\epsilon}{1-\zeta} (\mathrm{tr} \mathbf{C} - 3)$ & $-g_0$ & $0$ \\
EPTT                & $[0,2]$ & $-1$ & $1$ & $\exp\left[\frac{\epsilon}{1-\zeta}(\mathrm{tr} \mathbf{C} - 3)\right]$ & $-g_0$ & $0$ \\
\hline
\end{tabular}
\caption{Coefficients for various viscoelastic constitutive models in the generic tensorial formulation.}
\label{tab:constitutive-models}
\end{table}

The physical conformation tensor $\mathbf{C}^-$ is defined in the viscoelastic phase
$\Omega^-(t)$. The corresponding one-field polymeric stress is
\begin{equation}
    \boldsymbol{\tau}(\mathbf{x},t)
    = \chi(\mathbf{x},t)
      \frac{\eta^-}{(1-\zeta^-)\lambda^-}
      \left(h_0\mathbf{I} + h_1\mathbf{C}^- \right),
    \label{eq:phaseweightedstress}
\end{equation}
where $\lambda^-$ is the relaxation time, $\zeta^-$ is a material parameter, and $h_0$ and
$h_1$ are scalar functions (cf. \cref{tab:constitutive-models}). With no phase change, a
formal phase-specific conservative balance for $\mathbf{C}^-$ is
\begin{equation}\label{eq:cminusequation}
\frac{\partial (\chi\,\mathbf{C}^-)}{\partial t}
+ \nabla \cdot (\mathbf{v}\chi\mathbf{C}^-)
= \chi\left[
\mathbf{C}^-\cdot\mathbf{L}^\mathsf{T}
+ \mathbf{L}\cdot\mathbf{C}^-
+ \frac{1}{\lambda^-}\cvec{P}(\mathbf{C}^-)
\right],
\end{equation}
where $\mathbf{L}:=\mathbf{v}^\mathsf{T}-\zeta\mathbf{D}$ and
$\cvec{P}(\mathbf{C})=g_0\mathbf{I}+g_1\mathbf{C}+g_2\mathbf{C}^2$. The coefficients
$g_0$, $g_1$, and $g_2$ are isotropic scalar functions of $\mathbf{C}$ listed in
\cref{tab:constitutive-models}.

The present implementation does not use $\chi\mathbf{C}^-$ as the transported conformation
field. Instead, it introduces a positive-definite auxiliary extension $\mathbf{C}$ over the
complete computational domain, with $\mathbf{C}=\mathbf{C}^-$ in $\Omega^-(t)$, and solves
\begin{equation}\label{eq:cequation}
    \frac{\partial \mathbf{C}}{\partial t}
    + \nabla \cdot (\mathbf{v}\,\mathbf{C})
    = \mathbf{C}\cdot\mathbf{L}^\mathsf{T}
    + \mathbf{L}\cdot\mathbf{C}
    + \frac{1}{\lambda}\cvec{P}(\mathbf{C})
\end{equation}
throughout $\Omega$. The physical stress is then recovered by the phase weighting in
\cref{eq:phaseweightedstress}. Consequently, transformed representations such as
$\log\mathbf{C}$ are applied to the positive-definite auxiliary field $\mathbf{C}$, not to
$\chi\mathbf{C}^-$, which would be singular outside the viscoelastic phase. The numerical
consequences of transporting this extension over the full domain are discussed in
\cref{sec:numerics}.



\subsection{Stabilization}
The present work addresses the HWNP using a unified stabilization framework. It was developed for single-phase flow~\cite{Niethammer2018} and later extended to two-phase flow~\cite{Niethammer2019,Niethammer2025}. Here, we summarize its main components within the geometric VOF method.

DeboRheo provides the LCM and RCM stabilization methods. Here, we use LCM with $\mathbf{\Psi}=\ln\mathbf{C}$. The transposed velocity gradient is decomposed into the antisymmetric tensors $\mathbf{\Omega}$ and $\mathbf{N}$ and the symmetric tensor $\mathbf{B}$:
\begin{equation}
    \nabla\mathbf{v}^T = \mathbf{\Omega} + \mathbf{B} + \mathbf{N}\mathbf{C^{-1}}.
\label{eq:graddecomp}
\end{equation}
After applying this transformation and \cref{eq:graddecomp} to the auxiliary conformation equation \cref{eq:cequation}, the globally transported LCM equation is

\begin{align}
\label{eq:FCequation}
\frac{\partial \mathbf{\Psi}}{\partial t}
+ \nabla \cdot \bigl(\mathbf{v}\,\mathbf{\Psi}\bigr)
&= 2\,\mathbf{B} + \mathbf{\Omega}\cdot\mathbf{\Psi}
- \mathbf{\Psi}\cdot\mathbf{\Omega}
+ \frac{1}{\lambda}\,e^{-\mathbf{\Psi}}\cdot\cvec{P}\bigl(e^{\mathbf{\Psi}}\bigr).
\end{align}
In the present work, the natural logarithm conformation representation is found to be stable across various benchmark cases.

\section{Numerical method}
\label{sec:numerics}

We discretize the governing \cref{eq:volume-transport,eq:momentum-transport,eq:vof} and the auxiliary conformation equation \cref{eq:cequation}, together with constitutive equations for $\boldsymbol{\tau}$ \citep{Niethammer2025}, using the unstructured finite volume method and the unstructured geometrical Volume-of-Fluid (VOF) method. The formal phase-specific balance \cref{eq:cminusequation} is used to identify the consistency requirements at the interface but is not discretized directly.

The governing equations are solved in a segregated manner within OpenFOAM's
PIMPLE pressure--velocity coupling framework. The implementation optionally
enables the SIMPLEC correction~\citep{VanDormaal1984} by setting
\texttt{consistent true} in the PIMPLE controls. SIMPLEC is included because of
its effectiveness in viscoelastic flow simulations~\citep{Pimenta2016}.

\subsection{Volume conservation equation}

The volume conservation equation (\ref{eq:volume-transport}) is discretized by the unstructured collocated finite volume method \citep{Jasak1996} by integration over a fixed control-volume $\Omega_c$, i.e., 
\begin{equation}
    \int_{\Omega_c} \nabla\cdot\mathbf{v} \, dV =
    \int_{\partial \Omega_c} \mathbf{v} \cdot \mathbf{n} \, dS = 
     \sum_{f \in F_c} \int_{S_f} \mathbf{v} \cdot \mathbf{n} \, dS 
    =
    \sum_{f \in F_c} F_f + \epsilon_{\mathbf{v}}(h^2) = 0,
    \label{eq:voltranspdiscr}
\end{equation}
where $S_f \subset \partial \Omega_c$ denotes the (possibly non-planar) \emph{face} of the control volume, $h$ is the characteristic discretization length, and $\epsilon_{\mathbf{v}}(h^2)$ is the discretization error \citep{Juretic2010} resulting from the face-centered average of $\mathbf{v}$ associated with the centroid $\mathbf{x}_f$ of the face $S_f$, and 
\begin{equation}
    F_f := \mathbf{v}(\mathbf{x}_f) \cdot \mathbf{S}_f,
    \label{eq:volflux}
\end{equation}
is the volumetric flux.

\subsection{Volume-of-Fluid equation}
The volume-of-fluid equation (\ref{eq:vof}) is solved first, using the volumetric flux $F_f$  (\ref{eq:volflux}), by integration over the time-step $[t^n,t^{n+1}]$. Since the governing equations are solved in a segregated way, requiring outer iterations, we denote the outer iteration with the index - $o$.
\begin{equation}
    \alpha_c^{n+1} = \alpha_c^{n} - \frac{1}{|\Omega_c|}\int_{\partial\Omega_c}\int_{t^n}^{t^{n+1}} \chi(\mathbf{x}, t) \mathbf{v} \cdot\mathbf{n} dt dS.
    \label{eq:alphadiscrfromchi}
\end{equation}
The integral of the r.h.s. of \cref{eq:alphadiscrfromchi} we compute using the plicRDF-isoAdvector method \citep{Scheufler2019}, i.e. 
\begin{equation}
    \alpha_c^{n+1} = 
    \alpha_c^{n} - \frac{1}{|\Omega_c|} 
    \sum_{f\in F_c} V_f^\alpha
    \approx \alpha_c^{n} - \frac{1}{|\Omega_c|}\sum_{f\in F_c} \frac{0.5(F_f^{n+1} + F_f^n)}{|\mathbf{S_f}|} \int_{t^n}^{t^{n+1}} A_f(t) dt 
    \label{eq:alphaisoadvector}
\end{equation}
where $\int_{t^n}^{t^{n+1}} A_f(t) dt$ is computed geometrically \citep{Roenby2016}, to evaluate $V_f^\alpha$, the \emph{phase-specific volume} flux across $S_f$. 

Having solved for $\alpha_c^{n+1}$, $\rho_c^{n+1}$ and $\mu_c^{n+1}$ are updated by integrating \cref{eq:rhoindicator,eq:nuindicator} over $\Omega_c$, resulting in 
\begin{equation}
    \begin{aligned}
        \rho_c^{n+1} & = (\rho^ - - \rho^+) \alpha_c^{n+1} + \rho^+, \\
        \mu_c^{n+1} & = (\mu^- - \mu^+) \alpha_c^{n+1} + \mu^+.
    \end{aligned}
    \label{eq:rhonuc}
\end{equation}
The seemingly straightforward update of cell-centered density in \cref{eq:rhonuc} after solution of the VOF equation can potentially introduce inconsistency between mass (volume) and momentum conservation on discrete level, and it requires a careful choice of discretization schemes or an introduction of an auxiliary mass conservation equation for the one-field density \citep{Liu2023,Liu2024}. 

Additional information regarding the interaction between \cref{eq:alphaisoadvector,eq:rhonuc} and the other governing equations, which is relevant to the new method DeboRheo-v2.0, is provided in the sections below.

\subsection{Viscoelastic stress transport}

The auxiliary conformation equation \cref{eq:cequation} is solved before the momentum equation to compute the polymeric stress $\boldsymbol{\tau}$. A direct finite-volume discretization of the formal phase-specific balance \cref{eq:cminusequation}, however, presents major challenges. In principle, a phase-specific field $\phi=\chi\phi^-+(1-\chi)\phi^+$ requires averaging over phase-specific sub-volumes, i.e.,
\begin{equation}\label{eq:phipmh_emb}
|\Omega_c| \phi_c = |\Omega_c^-(t)|\,\phi^-(\mathbf{x}_c^-(t)) + |\Omega_c^+(t)|\,\phi^+(\mathbf{x}_c^+(t)) + O(h^2),
\end{equation}  
with $\mathbf{x}_c^\pm$ the centroids of the phase-specific cut volumes $\Omega_c^\pm$. In other words, a second-order accuracy is only maintained if phase-specific centroids and fluxes are introduced---essentially a cut-cell method---which is numerically fragile for cells with $\alpha_c\!\to 0,1$ or highly skewed cut volumes. Applying this reasoning to the viscoelastic tensor gives  
\begin{equation}\label{eq:cminusdiscr_emb}
\mathbf{C}_c \approx \alpha_c \mathbf{C}_c^-, 
\qquad
\partial_t \mathbf{C}_c \approx \partial_t(\alpha_c \mathbf{C}_c^-),
\end{equation}  
suggesting that simply transporting $\mathbf{C}$ everywhere might mimic the phase-specific form. But this makes the transport inconsistent with the geometrical VOF update of $\alpha_c$. This becomes apparent when discretizing the convective term. For the phase-specific flux,  
\begin{equation}\label{eq:cminusconv_emb}
\int_{\Omega_c} \nabla \cdot (\mathbf{v}\chi\mathbf{C}^-) \, dV 
= \sum_{f\in F_c} |S_f|\, (\mathbf{v}_f\cdot\mathbf{n}_f)\, \alpha_f \mathbf{C}_f^- + O(h^2),
\end{equation}  
with $\alpha_f$ the area fraction, defined as  
\begin{equation}\label{eq:alpha_face_emb}
\alpha_f = \frac{1}{|S_f|}\int_{S_f}\chi\,dS.
\end{equation}  
This shows that both $\alpha_f$ and $\mathbf{C}_f^-$ must be reconstructed in a way consistent with the geometrical VOF method. Standard unstructured FVM interpolation,  
\begin{equation}\label{eq:cminuslinear_emb}
\mathbf{C}_f^- = w_f \mathbf{C}_{Nf}^- + (1-w_f)\mathbf{C}_{Of}^-,
\end{equation}  
is inconsistent. Tourbier et al. \citep{Tourbier2024} have shown that for a constant scalar, consistency with isoAdvector--plicRDF is achieved instead by upwinding the $\alpha$--weighted product,  
\begin{equation}\label{eq:cminusalphaupw_emb}
\alpha_f \mathbf{C}_f^- = \alpha_{Uf} \mathbf{C}_{Uf}^-,
\end{equation}  
where the subscript $Uf$ is an index denoting the upwind cell with respect to the face $S_f$. This guarantees alignment with the geometric VOF update of $\alpha$, but at the cost of strong CFL restrictions when treated explicitly.  


State-of-the-art methods \citep{Figueiredo2016,LopezHerrara2019,Dritselis2022}, including our previous work \citep{Niethammer2025}, bypass these challenges by solving the one-field form of the conformation tensor equation for $\mathbf{C}$ everywhere, while applying phase-specific source terms. This approach, however, introduces a numerical inconsistency: $\mathbf{C}$ is advected across the interface with algebraic finite-volume schemes, causing numerical diffusion in the interface-normal direction, which is visualized in simulations as the ``leakage" of $\mathbf{C}$ (or its transformation) across the fluid interface. Additionally, the advection utilizes the full volumetric flux, rather than the phase-specific flux, resulting in a systematic overestimation and a flux of $\mathbf{C}$ across the interface.  

Tourbier et al.~\cite{Tourbier2024} provide an explicit approach for a scalar field containing a jump discontinuity, co-advected with volume fractions. This suggests a possible route to consistency for tensor quantities as well. However, for the non-constant conformation tensor, governed by complex transport equation, and discontinuous across $\Sigma(t)$, fundamental research is still required. For this reason, in the present work we adopt a pragmatic approach used by state-of-the-art methods. We solve the conformation tensor equation everywhere and recover the viscoelastic stress in the momentum equation by scaling with the indicator function $\chi$, or equivalently with the volume fraction $\alpha_c$.

\subsection{Momentum conservation equation}

This section covers the changes introduced to the momentum conservation equation relative to the first version of DeboRheo \citep{Niethammer2025}. In operator form, the finite-volume equation assembled by \texttt{UEqn.H} over a control volume \(\Omega_c\) is
\begin{align}
\frac{(\rho_c^o \mathbf{v}_c^o - \rho_c^n \mathbf{v}_c^n)|\Omega_c|}{\Delta t}
&+ \sum_{f\in F_{c}} \rho_f^o F_f^{o-1} \Bigl[ w_f\,\mathbf{v}_{\text{nb}(f)}^o + (1-w_f)\,\mathbf{v}_c^o \Bigr] \nonumber\\
&-\mathcal{D}^{\mathrm{imp}}_{\mu,c}(\mathbf{v}^o)
-\mathcal{D}^{\mathrm{imp}}_{\mathrm{BSD},c}(\mathbf{v}^o) \nonumber\\
&= -\,|\Omega_c|\,(\nabla p)^{o-1}_c - |\Omega_c|\,(\mathbf{g}\cdot\mathbf{x})_c\,(\nabla \rho)^o_c \nonumber\\[1mm]
&\quad +\, |\Omega_c|\,\sigma_c\,\kappa_c^o\,(\nabla \alpha)^o_c
+\mathcal{D}^{T,\mathrm{exp}}_{\mu,c}(\mathbf{v}^o)
+\mathcal{T}^{\mathrm{exp}}_{\widetilde{\alpha}\boldsymbol{\tau},c} \nonumber\\
&\quad -\mathcal{D}^{\mathrm{exp}}_{\mathrm{BSD},c}(\mathbf{v}^o),
\label{eq:full_momentum}
\end{align}

Here, $\mathcal{D}^{\mathrm{imp}}_{\mu,c}$ is the solvent-viscosity Laplacian assembled with the face viscosity defined in \cref{eq:mufinterpolated}, and $\mathcal{D}^{T,\mathrm{exp}}_{\mu,c}$ denotes the explicit variable-coefficient contribution corresponding in the implementation to $(\nabla\mathbf{v})\mathbin{\&}\nabla\mu_f$. The operator $\mathcal{T}^{\mathrm{exp}}_{\widetilde{\alpha}\boldsymbol{\tau},c}$ is the explicit divergence of the polymeric stress weighted by the bounded cell-centered volume fraction $\widetilde{\alpha}_c=\min(\max(\alpha_c,0),1)$. It does not use the geometrically reconstructed face area fraction. The BSD operators use the same coefficient $\gamma\widetilde{\alpha}_c\eta_c$ with opposite explicit and implicit contributions, so they cancel in the converged momentum equation.

\Cref{eq:full_momentum} is solved in the outer iteration (\emph{momentum predictor}) implicitly for $\mathbf{v}_c^o$, the cell-centered velocity field at outer iteration $o$. Terms marked ``exp'' are evaluated from the current or previous outer iterate, whereas the solvent and BSD Laplacians are assembled implicitly.

In \cref{eq:full_momentum}, the time derivative 
\(\partial_t(\rho \mathbf{v})\) is discretized as 
\(\bigl(\rho_c^o\,\mathbf{v}_c^o - \rho_c^n\,\mathbf{v}_c^n\bigr)/\Delta t\), 
where \(\rho_c^o\) is explicitly evaluated from \cref{eq:rhonuc}. 
This is done after the volume-of-fluid equation (\cref{eq:alphaisoadvector})
has been solved geometrically for \(\alpha_c^o\), maintaining consistency 
between mass (volume), phase-specific volume, and momentum 
\citep{Liu2023}, with the mass flux approximated from the fluxed 
phase-specific volume in \cref{eq:alphaisoadvector}.

The convective term $\nabla \cdot (\rho \mathbf{v}\otimes\mathbf{v})$ in \cref{eq:momentum-transport} is linearized; otherwise, its discretization would contain quadratic terms, resulting in a global nonlinear algebraic system. The velocity linearization uses the volumetric flux from the previous iteration, or the previous time step in the first outer iteration, $F_f^{o-1}$. The mass flux $\rho_f^o F_f^{o-1}$ is reconstructed from the phase-specific volume integrated in \cref{eq:alphaisoadvector}. Without this construction, discrete volume, phase-volume, and momentum transport can become inconsistent. A detailed analysis for geometrical VOF methods at high density ratios is provided by Liu et al.~\cite{Liu2023}. In the first-order upwind cases, the Euler temporal discretization and the approximation $\rho_f^o F_f^{o-1}\approx (\rho^- - \rho^+)V_f^\alpha/\Delta t$ follow that consistency argument. The benchmark dictionaries use first-order upwind interpolation for the shear-driven, Newtonian, and Prieto cases, and \texttt{limitedLinearV 1} for the Zainali and axisymmetric Esposito cases.

The face-centered velocity \(\mathbf{v}_f^o\) in $\sum_{f \in F_c} \rho_f^o F_f^{o-1} \mathbf{v}_f^o$ is expressed in \cref{eq:full_momentum} as a linear combination of the neighboring-cell and owner-cell velocities using the interpolation weight \(w_f\). In the unstructured FVM \citep{Jasak1996}, $w_f$ depends on the convective interpolation scheme selected for the individual benchmark.

\subsection{Interfacial treatment of solvent viscosity}
\label{sec:interfacial-viscosity}

The solvent-viscosity contribution is assembled from an implicit Laplacian and an explicit variable-coefficient term. Both require a face-centered viscosity after the geometrical VOF update in \cref{eq:alphaisoadvector}. DeboRheo v1.0 used the geometrically reconstructed face area fraction to compute $\mu_f=(\mu^- - \mu^+)\alpha_f+\mu^+$, with
\begin{equation}
    \alpha_f := \frac{1}{|S_f|}\int_{S_f} \chi(\mathbf{x}, t) dS,
    \label{eq:alphaf}
\end{equation}
where $\alpha_f$ is the geometric face-area fraction. DeboRheo v2.0 instead uses the standard finite-volume interpolation operator $\mathcal{I}_f$ separately for the cell-centered volume fraction and each phase viscosity:
\begin{align}
    \widetilde{\alpha}^{I,o}_f
    &:= \min\!\left(\max\!\left(\mathcal{I}_f(\alpha_c^o),0\right),1\right), \nonumber\\
    \mu_f^o
    &= \widetilde{\alpha}^{I,o}_f\,\mathcal{I}_f(\eta_{S,c}^o)
    + \left(1-\widetilde{\alpha}^{I,o}_f\right)\mathcal{I}_f(\eta_{2,c}^o).
    \label{eq:mufinterpolated}
\end{align}
Here, $\eta_{S,c}$ and $\eta_{2,c}$ denote the solvent viscosity of phase 1 and the viscosity of phase 2, respectively, matching the implementation variables \texttt{etaS} and \texttt{eta2()}. For constant phase viscosities, \cref{eq:mufinterpolated} is equivalent to linearly interpolating the cell-centered mixture viscosity. For spatially varying phase viscosities, the separate interpolations in \cref{eq:mufinterpolated} are the operative definition.

The interpolated fraction $\widetilde{\alpha}^{I}_f$ is distinct from the geometric area fraction $\alpha_f$ used by VOF advection. It produces a stencil-scale, mesh-dependent transition in the viscous coefficient while leaving the geometrically reconstructed interface sharp. The width of this transition depends on the mesh and interpolation stencil and is therefore not asserted to be exactly one cell. This regularization substantially improves robustness for thin lamellae and filaments. DeboRheo v1.0 cannot complete this thin-structure regime; consequently, no quantitative v1.0 comparison curve is reported.

 The cell-centered source terms on the r.h.s. of the momentum equation (\ref{eq:full_momentum}), i.e., the pressure gradient, the hydrostatic term, and the surface tension force, are reconstructed from face-centered values using linear least squares reconstruction described in detail in \citep{Tolle2020},
\begin{equation}
  \boldsymbol{\phi}^R_c \approx \mathcal{R}(F^\phi_f) = \left[ \sum_{\tilde{f}} \mathbf{S}_{\tilde{f}} \mathbf{S}_{\tilde{f}}  \right]^{-1} \cdot \sum_f \mathbf{\hat{S}}_f F^\phi_f = \mathbf{S}_c^{-1} \cdot  \sum_f \mathbf{\hat{S}}_f F^\phi_f + \boldsymbol{\epsilon}^R_c(h^p),
  \label{eq:reconstruct}
\end{equation}
with 
\begin{equation}
    F^\phi_f := \boldsymbol{\phi}_f \cdot \mathbf{S}_f.
\end{equation}

Terms related to the fluid interface, such as $(\nabla \rho)_c^o$, $\kappa_c^o$ and $(\nabla \alpha)_c^o$, are evaluated explicitly after solving the volume-of-fluid equation in \cref{eq:alphaisoadvector}.

\subsection{Pressure Poisson equation}

We consider the discretized and linearized momentum \cref{eq:full_momentum} written in global operator form as
\begin{equation}
\mathbf{A}^{\,o}\,\mathbf{v}^{\,o+1}
=
-\,\mathbf{G}\,\mathbf{p}^{\,o}
+
\mathbf{S}(\mathbf{v}^{\,o}),
\label{eq:mom_global}
\end{equation}
where \(\mathbf{A}^{\,o}\) is the full sparse momentum operator assembled at outer iteration
\(o\), \(\mathbf{G}\) is the discrete gradient operator, and
\(\mathbf{S}(\mathbf{v}^{\,o})\) collects all explicit source contributions, including old-time
terms and explicitly treated nonlinearities, evaluated from the previous outer iterate
\(\mathbf{v}^{\,o}\).
The operator \(\mathbf{A}^{\,o}\) depends on the volumetric fluxes \(F_f^{\,o}\) through the
linearization of convection and diffusion terms, and is therefore updated once per outer
iteration.
In the outer SIMPLE iteration, \cref{eq:mom_global} (i.e., \cref{eq:full_momentum}) is solved
using a Krylov-type linear solver.
Although the inverse \((\mathbf{A}^{\,o})^{-1}\) is never formed explicitly, the effect of this
solve can be written formally as
\begin{equation}
\mathbf{v}^{\,o+1}
=
-\,(\mathbf{A}^{\,o})^{-1}\mathbf{G}\,\mathbf{p}^{\,o}
+
(\mathbf{A}^{\,o})^{-1}\mathbf{S}(\mathbf{v}^{\,o}),
\label{eq:mom_formal_inverse}
\end{equation}
which shows that the predicted velocity field \(\mathbf{v}^{\,o+1}\) contains the full neighbour
coupling encoded in \(\mathbf{A}^{\,o}\).
This neighbour coupling is directly visible in \cref{eq:full_momentum}, for example through the
discretized convection and diffusion operators.

To derive a pressure Poisson equation that can be assembled efficiently, the momentum operator
is split into diagonal and off-diagonal parts,
\begin{equation}
\mathbf{A}^{\,o} = \mathbf{A}_c^{\,o} + \mathbf{A}_n^{\,o}.
\label{eq:A_split}
\end{equation}
For every cell "c", \cref{eq:mom_global} (see also \cref{eq:full_momentum}) can be written as
\begin{equation}
a_c^{\,o}\,\mathbf{v}_c^{\,o+1}
+
\sum_n a_n^{\,o}\,\mathbf{v}_n^{\,o+1}
=
-\,(\nabla \mathbf{p})_c^{\,o}
+
\mathbf{S}_c(\mathbf{v}^{\,o}),
\label{eq:mom_cell}
\end{equation}
where $a_c$ is the diagonal coefficient in $\mathbf{A}$, and $a_n$ are off-diagonal coefficients in $\mathbf{A}$ in the $c$-th matrix row.

By isolating the diagonal contribution,
\begin{equation}
a_c^{\,o}\,\mathbf{v}_c^{\,o+1}
=
-\,(\nabla \mathbf{p})_c^{\,o}
+
\mathbf{H}_c^{\,o},
\label{eq:mom_H_form}
\end{equation}
where the auxiliary field \(\mathbf{H}\) is defined as
\begin{equation}
\mathbf{H}_c^{\,o}
:=
- \sum_n a_n^{\,o}\,\mathbf{v}_n^{\,o}
+
\mathbf{S}_c(\mathbf{v}^{\,o}).
\label{eq:H_definition}
\end{equation}
Here, all neighbour velocities appearing in \(\mathbf{H}\) are evaluated from the previous outer
iterate \(\mathbf{v}^{\,o}\).
This corresponds to the standard segregated linearization step: all unknown quantities at
iteration \(o+1\) are retained on the left-hand side of
\cref{eq:mom_cell}, while all quantities known from the previous outer iteration are moved to
the right-hand side and incorporated into \(\mathbf{H}^{\,o}\).

Dividing \cref{eq:mom_H_form} by the diagonal coefficient \(a_c^{\,o}\) yields
\begin{equation}
\mathbf{v}_c^{o+1}
=
-\,(\mathbf{A}_c^{\,o})^{-1}\mathbf{G}\,\mathbf{p}
+
(\mathbf{A}_c^{\,o})^{-1}\mathbf{H}^{\,o},
\label{eq:v_diag_inverse}
\end{equation}
which replaces the action of the full inverse \((\mathbf{A}^{\,o})^{-1}\) in
\cref{eq:mom_formal_inverse} by the diagonal inverse \((\mathbf{A}_c^{\,o})^{-1}\).
This constitutes the central approximation introduced in SIMPLE- and PISO-type algorithms for
the pressure--velocity coupling.

Discrete incompressibility is enforced by requiring the discrete divergence of the velocity
field, i.e., the sum of volumetric face fluxes, to vanish,
\begin{equation}
\sum_f F_f^{\,i} = 0,
\label{eq:divflux}
\end{equation}
where the superscript \(i\) denotes the pressure-correction (inner) iteration carried out with
the momentum operator \(\mathbf{A}^{\,o}\) held fixed.
Using \cref{eq:v_diag_inverse}, the face flux at inner iteration \(i\) is constructed via a
Rhie--Chow-type interpolation,
\begin{equation}
F_f^{\,i}
=
\mathcal I_f\!\left((\mathbf{A}_c^{\,o})^{-1}\mathbf{H}^{\,o}\right)\cdot \mathbf{S}_f
-
\mathcal I_f\!\left((\mathbf{A}_c^{\,o})^{-1}\right)
\left(\mathbf{G}_f \mathbf{p}^{\,i}\right)\cdot \mathbf{S}_f,
\label{eq:RC_flux}
\end{equation}
where \(\mathcal I_f\) denotes interpolation from cell centres to face centres and
\(\mathbf{G}_f \mathbf{p}^{\,i}\) is the discrete face-normal pressure gradient.
Substitution of \cref{eq:RC_flux} into \cref{eq:divflux} yields the discrete pressure Poisson
equation (\cref{eq:pressure_poisson} in alternative form),
\begin{equation}
\sum_f
\left((\mathbf{A}_c^{\,o})^{-1}\right)_f
(\nabla \mathbf{p}^{\,i})_f \cdot \mathbf{S}_f
=
\sum_f
\mathcal I_f\!\left((\mathbf{A}_c^{\,o})^{-1}\mathbf{H}^{\,i}\right)\cdot \mathbf{S}_f.
\label{eq:pressure_poisson}
\end{equation}

In this formulation, the coefficients \((\mathbf{A}_c^{\,o})^{-1}\) are evaluated once per outer iteration and held fixed during the inner
pressure solution. In each inner iteration the auxiliary field
\(\mathbf{H}^{\,i}\) is evaluated using the velocity $v^i$ available at that inner iteration, given by \cref{eq:v_diag_inverse}. 
After solving \cref{eq:pressure_poisson} for \(\mathbf{p}^{\,i}\), the cell centered velocity is updated to $v^{i+1}$ by \cref{eq:v_diag_inverse} and face fluxes
\(F_f^{\,i}\) are updated using \cref{eq:RC_flux}, yielding a divergence-free flux field by
construction.
These updated fluxes are then used to reassemble the momentum operator
\(\mathbf{A}^{\,o+1}\) in the next outer iteration, thereby recovering the full neighbour
coupling through the momentum predictor step while retaining a computationally tractable
pressure equation.

\subsection{The SIMPLEC Algorithm}

Ideally, at outer iteration $o+1$, we would like the discretized and linearized momentum equation to be satisfied in the form
\begin{equation}
\mathbf{A}^{\,o}\,\mathbf{v}^{\,o+1}
=
-\,\mathbf{G}\,\mathbf{p}^{\,o+1}
+
\mathbf{S}(\mathbf{v}^{\,o}),
\label{eq:mom_target}
\end{equation}
where $\mathbf{A}^{\,o}$ denotes the momentum operator assembled at outer iteration $o$, based on the current linearization of the convective, diffusive, and additional stress terms, and $\mathbf{S}(\mathbf{v}^{\,o})$ collects all explicit contributions evaluated from the previous outer iterate.

However, when solving the momentum predictor at outer iteration $o+1$, the updated pressure $\mathbf{p}^{\,o+1}$ is not yet available. Instead, the momentum equation is solved using the pressure field from the previous outer iteration, yielding the predicted velocity $\mathbf{v}^{\,*}$,
\begin{equation}
\mathbf{A}^{\,o}\,\mathbf{v}^{\,*}
=
-\,\mathbf{G}\,\mathbf{p}^{\,o}
+
\mathbf{S}(\mathbf{v}^{\,o}),
\label{eq:mom_predictor}
\end{equation}
where $\mathbf{v}^{\,*}$ denotes the pressure-lagged velocity prediction at outer iteration $o+1$. The predicted velocity $\mathbf{v}^{\,*}$ is therefore neither equal to the velocity from the previous outer iteration $\mathbf{v}^{\,o}$, nor to the final velocity $\mathbf{v}^{\,o+1}$ that would satisfy the fully pressure-consistent momentum balance~\cref{eq:mom_target}.

Since the operators appearing in \cref{eq:mom_target} and \cref{eq:mom_predictor} are linear, these two equations can be subtracted to obtain a correction equation. Introducing the velocity and pressure corrections
\begin{equation}
\mathbf{v}' = \mathbf{v}^{\,o+1} - \mathbf{v}^{\,*},
\qquad
\mathbf{p}' = \mathbf{p}^{\,o+1} - \mathbf{p}^{\,o},
\label{eq:corrections_def}
\end{equation}
leads to the exact correction relation
\begin{equation}
\mathbf{A}^{\,o}\,\mathbf{v}' = -\,\mathbf{G}\,\mathbf{p}'.
\label{eq:correction_exact}
\end{equation}
Equation~\eqref{eq:correction_exact} shows that, under a frozen momentum linearization, the velocity correction is governed by the inverse of the full momentum operator $\mathbf{A}^{\,o}$. Formally, this relation can be written as $\mathbf{v}' = -(\mathbf{A}^{\,o})^{-1}\mathbf{G}\mathbf{p}'$. However, the explicit construction or application of $(\mathbf{A}^{\,o})^{-1}$ is computationally intractable within a segregated pressure--velocity coupling framework.

The PISO and SIMPLE families of algorithms therefore approximate the velocity--pressure correction relation by retaining only the diagonal part of the momentum operator when assembling the pressure (correction) Poisson equation. Denoting the diagonal of $\mathbf{A}^{\,o}$ by $\mathbf{A}_c^{\,o}$, the velocity correction is approximated as
\begin{equation}
\mathbf{v}' \approx -\,(\mathbf{A}_c^{\,o})^{-1}\mathbf{G}\,\mathbf{p}'.
\label{eq:simple_correction}
\end{equation}
This approximation neglects the contribution of the off-diagonal momentum couplings to the velocity response, and systematically underestimates the effect of pressure corrections in strongly coupled momentum systems.

The SIMPLEC approach seeks to incorporate the influence of the off-diagonal operator $\mathbf{A}_n^{\,o}$, defined through the splitting
\begin{equation}
\mathbf{A}^{\,o} = \mathbf{A}_c^{\,o} + \mathbf{A}_n^{\,o},
\end{equation}
into the approximation of the inverse momentum response. Writing the correction (eq. (\ref{eq:correction_exact})) at the cell level yields
\begin{equation}
a_c^{\,o}\,\mathbf{v}_c'
+
\sum_n a_n^{\,o}\,\mathbf{v}_n'
=
-\,(\nabla \mathbf{p}')_c,
\end{equation}
where $a_c^{\,o}$ and $a_n^{\,o}$ denote the diagonal and off-diagonal coefficients of the momentum operator, respectively. The SIMPLEC approximation assumes that, within a single outer iteration, the pressure-induced velocity correction varies smoothly across neighbouring cells, such that the neighbour corrections can be approximated locally as
\begin{equation}
\mathbf{v}_n' \approx \mathbf{v}_c'.
\end{equation}
Under this assumption, the correction equation reduces to
\begin{equation}
\left(a_c^{\,o} - \sum_n a_n^{\,o}\right)\mathbf{v}_c'
\;\approx\;
-\,(\nabla \mathbf{p}')_c,
\end{equation}
or equivalently
\begin{equation}
\mathbf{v}_c'
\;\approx\;
-\,(\mathbf{A}_{t,c}^{\,o})^{-1}\mathbf{G}\,\mathbf{p}',
\end{equation}
where
\begin{equation}
\mathbf{A}_{t,c}^{\,o}
=
\operatorname{diag}\!\left(a_c^{\,o} - \sum_n a_n^{\,o}\right)
\label{eq:Asimplec}
\end{equation}
defines the SIMPLEC approximation of the inverse momentum operator appearing in~\cref{eq:correction_exact}.

Compared to the standard SIMPLE approximation~\cref{eq:simple_correction}, the modified inverse $(\mathbf{A}_{t,c}^{\,o})^{-1}$ provides a stronger and more consistent coupling between pressure and velocity corrections by accounting, in an aggregated manner, for neighbour momentum couplings. This enhanced pressure--velocity coupling has been shown to improve robustness and convergence in strongly coupled flow regimes, including two-phase viscoelastic flows~\citep{Alves2021}, and is therefore available in the \texttt{deboRheo}~v2.0 framework through the \texttt{consistent true} PIMPLE setting.

\subsection{Time-Step Selection}

The time step is determined by two stability constraints. The first constraint is based on the maximum Courant number criterion,

\begin{equation}
    \Delta t_{\mathrm{Co}} =
    \frac{\mathrm{Co}_{\max}\,\Delta x}{|\mathbf{v}|},
\end{equation}
where $\Delta x$ is the minimum characteristic cell length and  $\mathrm{Co}_{\max}=0.01$~\cite{Esposito2023}. The second constraint is based on the characteristic capillary-wave time scale,
\begin{equation}
    \Delta t_{\sigma}
    =
    \sqrt{
    \frac{0.5(\rho^{+}+\rho^{-})\,\Delta x^{3}}
    {\sigma}}.
\end{equation}
This constraint ensures that capillary-wave propagation at the fluid interface is adequately resolved and is particularly important for capillary-dominated flows. The time step employed in the simulations is given by
\begin{equation}
    \Delta t
    =
    \min\left(
    c_{1}\Delta  t_{\mathrm{Co}},
    c_{2}\Delta t_{\sigma}
    \right),
\end{equation}
where $c_{1}$ and $c_{2}$ are safety factors set to $1$ and $0.1$, respectively.

\section{Software design}
\label{sec:software-design}

\subsection{Scope}
The DeboRheo framework provides a software implementation of the numerical solution
strategy for viscoelastic flows introduced in Sections~\ref{sec:model}
and~\ref{sec:numerics}. It extends incompressible single- and two-phase flow solvers by
incorporating viscoelastic constitutive equations and their coupling to the momentum
equation. For two-phase configurations, the framework reuses the geometric
Volume-of-Fluid (VOF) methodology of the \texttt{TwoPhaseFlow} OpenFOAM module \citep{TwoPhaseFlow} and augments it with a
phase-dependent polymeric stress formulation. The DeboRheo 2.0 has been tested and verified to work with OpenFOAM v2512.

At each time step, a transport equation is solved for a tensorial state variable associated
with the viscoelastic model. The polymeric stress is reconstructed from this state and its
divergence is added explicitly to the right-hand side of the momentum equation. The role of
the DeboRheo framework is to organize this procedure at the solver level. Constitutive
models, numerical representations of the viscoelastic state, and optional stabilization
mechanisms are implemented as separate components that can be selected at run time. This
structure allows the numerical workflow defined by the governing equations to be preserved
while supporting different constitutive closures and state-variable formulations within a
single solver framework.

\subsection{Solver file organization}

The \texttt{interIsoDeboRheoFoam} solver follows the standard OpenFOAM design, in which
the simulation workflow is coordinated by a main driver file, while field initialization
and equation assembly are implemented in dedicated include files. In addition to its
solver-specific implementation, \texttt{interIsoDeboRheoFoam} reuses components of the
\texttt{TwoPhaseFlow} module for geometric Volume-of-Fluid (VOF) interface capturing and
time-step control. This modular organization separates solver control from equation
assembly and promotes code reuse across different multiphase flow solvers.

The solver-specific directory is organized as follows:

\begin{lstlisting}
interIsoDeboRheoFoam/
|-- interIsoDeboRheoFoam.C   // main solver driver and time loop
|-- createFields.H           // field creation and model initialization
|-- UEqn.H                   // momentum equation assembly
`-- pEqn.H                   // pressure correction equation
\end{lstlisting}

The geometric VOF routines and adaptive time-step control are provided by the
\texttt{TwoPhaseFlow} module, which contributes functionality such as phase-fraction
transport, interface reconstruction, and time-step adjustment through reusable include
files (e.g., \texttt{alphaEqn.H}, \texttt{createIsoAdvection.H}, and
\texttt{setDeltaT.H}).

The main solver file coordinates the simulation time loop and invokes the individual
equation files in a prescribed sequence. Field creation and model initialization are
performed in \texttt{createFields.H}, while the momentum and pressure equations are
assembled in \texttt{UEqn.H} and \texttt{pEqn.H}, respectively. The phase-fraction
transport and interface reconstruction are delegated to the \texttt{TwoPhaseFlow}
module, which supplies the geometric VOF implementation used to compute the interface
geometry and phase distribution required for evaluating phase-dependent material
properties and viscoelastic stress coupling. This separation mirrors the numerical
structure of the solver while maintaining a clear distinction between solver-specific
logic, reusable multiphase-flow functionality, and the DeboRheo rheology libraries.

\subsection{Numerical solver design}
The solver \texttt{interIsoDeboRheoFoam} realizes the numerical solution strategy described
in Section~\ref{sec:numerics} within the segregated PIMPLE pressure--velocity coupling
framework. The PIMPLE dictionary enables consistent SIMPLE by setting the keyword
\texttt{consistent} to \texttt{true}. The corresponding \texttt{pimple.consistent()} branch
then activates the SIMPLEC correction. In two-phase configurations, the solver reuses the geometric Volume-of-Fluid
(VOF) method of the \texttt{TwoPhaseFlow} module to advance the phase fraction and reconstruct
the interface. The geometric face area fractions are passed through the mixture-model
interface, whereas the active log-conformation implementation weights the polymeric stress
with a bounded cell-centered volume fraction.

At each time step, the viscoelastic state is advanced by solving a transport equation
associated with the selected constitutive model and numerical representation, yielding an
updated polymeric stress tensor. The momentum equation is then assembled according to
\cref{eq:full_momentum,eq:mom_global}. The mixture model supplies the phase-weighted polymeric
stress divergence and, when BSD is active, the paired explicit and implicit diffusion terms:
\begin{lstlisting}[language=C++]
fvVectorMatrix UEqn
(
    fvm::ddt(rho, U)
  + fvm::div(rhoPhi, U)
  + MRF.DDt(rho, U)
  - (fvc::grad(U) & fvc::grad(mixture.muf()))
  - fvm::laplacian(mixture.muf(), U)
  - mixture.divTau(areaFractionRecon.areaFractions(), U)
 ==
    fvOptions(rho, U)
);

UEqn.relax();
fvOptions.constrain(UEqn);
\end{lstlisting}
Here, \texttt{mixture.divTau} returns a finite-volume operator containing the explicitly
evaluated divergence of the cell-volume-fraction-weighted polymeric stress. With BSD enabled,
it also contains an explicitly subtracted diffusion term and the matching implicit Laplacian;
these contributions cancel in the converged equation and therefore do not constitute net
artificial diffusion.

The new evaluation of $\mu_f$ in DeboRheo v2.0 is implemented as 
\begin{lstlisting}[language=C++]
const surfaceScalarField alpha1f
(
    min(max(fvc::interpolate(alpha1_), scalar(0)), scalar(1))
);

return tmp<surfaceScalarField>::New
(
    "muf",
    alpha1f*fvc::interpolate(etaS)
  + (scalar(1) - alpha1f)*fvc::interpolate(eta2())
);
\end{lstlisting}

The evaluation of $\mu_f$ in this function separately interpolates the cell-centered volume
fraction and the two phase viscosities. It is therefore distinct from the geometrically
reconstructed face area fraction used by the VOF advection algorithm in v1.0. This deliberate
distinction implements the interfacial viscosity regularization described in
\cref{sec:interfacial-viscosity}.

The velocity predictor is formed from the discretized momentum equation and used to enforce
the incompressibility constraint through the pressure Poisson equation 
(see \cref{eq:pressure_poisson}), formulated in terms of the modified pressure \(p_{rgh}\). 
The latter incorporates the hydrostatic pressure contribution, ($p_{rgh}=p-\rho gh$), thereby
accounting for fluid density variations in the pressure correction step. The diagonal inverse
$r_{AU}=1/A$ is always formed. If \texttt{pimple.consistent()} is true, the SIMPLEC inverse
$r_{AtU}=1/(1/r_{AU}-H_1)$ (see \cref{eq:Asimplec}) and its associated predictor corrections
are activated. Otherwise, $r_{AtU}$ remains equal to $r_{AU}$ and the standard pressure
correction is used. The corresponding predictor flux $\phi_{H/A}$ is assembled and used on
the right-hand side of the Poisson equation:
\begin{lstlisting}[language=C++]
surfaceScalarField rAUf("rAUf", fvc::interpolate(rAU()));
volVectorField HbyA(constrainHbyA(rAU()*UEqn.H(), U, p_rgh));
surfaceScalarField phiHbyA
(
    "phiHbyA",
    fvc::flux(HbyA)
  + MRF.zeroFilter
    (
        fvc::interpolate(rho*rAU())
      * fvc::ddtCorr(U, phi, Uf)
    )
);

tmp<volScalarField> rAtU(rAU);
if (pimple.consistent())
{
    rAtU = 1.0/(1.0/rAU() - UEqn.H1());
    phiHbyA += fvc::interpolate(rAtU() - rAU())
        *fvc::snGrad(p_rgh)*mesh.magSf();
    HbyA -= (rAU() - rAtU())
        *fvc::reconstruct(fvc::snGrad(p_rgh)*mesh.magSf());
}

while (pimple.correctNonOrthogonal())
{
    fvScalarMatrix p_rghEqn
    (
        fvm::laplacian(rAtU(), p_rgh) == fvc::div(phiHbyA)
    );

    p_rghEqn.solve
    (
        mesh.solver(p_rgh.select(pimple.finalInnerIter()))
    );

    if (pimple.finalNonOrthogonalIter())
    {
        phi = phiHbyA - p_rghEqn.flux();
        U = HbyA
          + rAU()*fvc::reconstruct
            (
                (phig - p_rghEqn.flux())/rAUf
            );
    }
}
\end{lstlisting}
Here, \texttt{p\_rghEqn} represents the pressure Poisson equation, and the corrected velocity
field is obtained by projection using the computed pressure field. The pressure equation is
not modified explicitly by the viscoelastic model. Instead, the polymeric stress contribution
enters through the momentum predictor and is therefore contained in the \texttt{UEqn.H()}
operator. In the case dictionaries, SIMPLEC must be selected with \texttt{consistent true}; the
keyword \texttt{SIMPLEC true} is not queried by this implementation.

\subsection{Library organization}

The DeboRheo functionality is implemented as a collection of libraries that extend the
transport and mixture models used by the flow solvers. The library structure reflects the
software decomposition of the viscoelastic framework into constitutive modeling,
constitutive-variable representation, and transport-model coupling. This modular
organization allows the solver to interact with the viscoelastic subsystem through a small
set of well-defined interfaces, while encapsulating model-specific implementations and
representation-specific numerical formulations within the library layer.

The relevant library directories are organized as follows:

\begin{lstlisting}
src/
|-- rheology/
|   |-- rheologicalModel/          // model interface
|   |-- rheologicalModels/         // constitutive models
|   |-- representation/            // state representations
|   |   |-- representation/        // base classes
|   |   |-- genericRepresentation/ // tensor transforms
|   |   |-- stressRepresentation/  // stress form
|   |   |-- logConformationRepresentation/
|   |   |-- rootConformationRepresentation/
|   |   `-- multiMode/             // multimode support
|   `-- diagonalize/               // tensor diagonalization utilities
|
`-- twoPhaseTransportModels/
    `-- ...                        // phase coupling
\end{lstlisting}

The \texttt{rheologicalModel} layer provides the abstract interface for constitutive
models, performs runtime selection of the requested rheological model, and exposes a
common application programming interface (API) through which the solver accesses
viscoelastic stresses and constitutive updates. Individual constitutive models implement
the corresponding constitutive equations together with their material-specific parameters.

The \texttt{representation} layer defines the constitutive state variable and provides
the mathematical transformations and common infrastructure associated with different
formulations of the constitutive equations. These include stress-based, log-conformation,
and root-conformation representations, together with shared functionality for tensor
transformations and multimode constitutive formulations.

The \texttt{twoPhaseTransportModels} layer provides the coupling between the viscoelastic
subsystem and the momentum equations by incorporating phase-dependent transport
properties and viscoelastic stress contributions. Through this modular architecture, the
solver remains independent of the specific constitutive model and state-variable
representation, interacting only with the abstract rheological interface, while the
underlying library encapsulates the implementation details of constitutive modeling and
numerical representation.

\subsection*{Viscoelastic library implementation}

In the solver, the viscoelastic contribution is evaluated before assembly of the momentum
equation. At each time step, the mixture model advances the viscoelastic state by solving
the constitutive transport equation associated with the selected model and numerical
representation,
\begin{lstlisting}[language=C++]
mixture.correctPolymer(areaFractionRecon.areaFractions(), false);
\end{lstlisting}
In the two-phase mixture, this call delegates the computation to the active numerical
representation,
\begin{lstlisting}[language=C++]
void incompressibleTwoPhaseMixtureDeboRheo::correctPolymer
(
    const surfaceScalarField& alpha1f,
    const bool& lastIter
)
{
    representationPtr_->correct(alpha1f, lastIter);
}
\end{lstlisting}
The face-area-fraction argument is retained in this interface for specific representations.
In the active log-conformation implementation shown below, the transported
state is advanced over the complete domain and the stress divergence is weighted using the
bounded cell-centered phase fraction rather than \texttt{alpha1f}.

\paragraph{Representations of the viscoelastic state}

The central abstraction used in the DeboRheo library is the \emph{representation}. A
representation defines the tensorial state variable that is transported in time and the
transformations required to recover the physical conformation tensor and polymeric stress.
Rather than solving directly for the conformation tensor $\boldsymbol{C}$, the library
solves for a generic tensor field $\boldsymbol{A}$, which is related to $\boldsymbol{C}$ by
a transformation
\begin{equation}
\boldsymbol{A} = \mathcal{T}(\boldsymbol{C}).
\end{equation}
Typical choices include $\boldsymbol{A}=\boldsymbol{C}$, $\boldsymbol{A}=\log\boldsymbol{C}$,
or $\boldsymbol{A}=\boldsymbol{C}^{1/k}$, with, e.g., $k=2$ or $k=3$. These transformations improve numerical stability
and preserve physical properties such as positive definiteness of the conformation tensor.

Each representation implements the same conceptual workflow: the constitutive law updates
model-specific quantities, the transport equation for $\boldsymbol{A}$ is solved, and the
polymeric stress is reconstructed. An example implementation using an uncorrected
log-conformation representation is shown below,
\begin{lstlisting}[language=C++]
void logConformationUncorrected::correct
(
    const surfaceScalarField& alpha1f,
    const bool& lastIter
)
{
    law_->correctPolymer();
    this->correctTwoPhase(alpha1f, lastIter);
    law_->correctThermo();
}
\end{lstlisting}

The transport equation for $\boldsymbol{A}$ is assembled and solved in the representation,
including advection, deformation, and relaxation terms,
\begin{lstlisting}[language=C++]
tmp<fvMatrix<symmTensor> > AEqn
(
    fvm::ddt(A_)
  + fvm::div(phi(), A_, "div(phi,tau)")
 ==
    rhst()
);
\end{lstlisting}

\paragraph{Stress reconstruction and divergence}
After solving for the conformation tensor $\boldsymbol{A}$, the polymeric stress
$\boldsymbol{\tau}$ is reconstructed according to the selected constitutive
representation. The divergence of the polymeric stress required in the momentum equation
is evaluated through the mixture-model interface, which forwards the request to the active
stress representation,
\begin{lstlisting}[language=C++]
tmp<fvVectorMatrix>
incompressibleTwoPhaseMixtureDeboRheo::divTau
(
    const surfaceScalarField& alpha1f,
    volVectorField& U
)
{
    return representationPtr_->divTau(alpha1f, U);
}
\end{lstlisting}

For representations based on transformed conformation tensors, the polymeric stress
divergence is implemented in the representation-specific \texttt{divTau} routine. This
routine forms the coefficient \texttt{etaPEff\_} and delegates assembly to the templated
helper \texttt{divTauT}. The latter weights the polymeric stress with the bounded
cell-centered phase fraction \texttt{limitedAlpha1}. When \texttt{alpha\_} is nonzero, the
BSD formulation subtracts an explicit diffusion operator and adds the identical operator
implicitly. Thus BSD changes the algebraic splitting and convergence behavior without adding
net diffusion to the converged momentum equation. When \texttt{alpha\_} is zero, only the
phase-weighted polymeric stress divergence remains.

\begin{lstlisting}[language=C++]
Foam::tmp<Foam::fvVectorMatrix>
Foam::logConformationUncorrected::divTau
(
    const surfaceScalarField& alpha1f, volVectorField& U
)
{
    if (alpha_ > SMALL)
    {
        etaPEff_ = alpha_*law_->etaP();
    }

    return divTauT<volScalarField>(alpha1f, U);
}

tmp<fvVectorMatrix>
logConformationUncorrected::divTauT
(
    const surfaceScalarField& alpha1f,
    volVectorField& U
) const
{
    volScalarField limitedAlpha1 =
        min(max(alpha1(), scalar(0)), scalar(1));

    if (alpha_ > SMALL)
    {
        return
            fvc::div(tau_*limitedAlpha1)
          - fvc::div((etaPEff_*limitedAlpha1)*fvc::grad(U))
          + fvm::laplacian((etaPEff_*limitedAlpha1), U);
    }

    return fvc::div(tau_*limitedAlpha1);
}
\end{lstlisting}

Overall, the library design isolates constitutive modelling, numerical representation, and solver coupling. The solver receives only the assembled contribution of the polymeric stress to the momentum equation, while all details related to viscoelastic transport, stabilization, and tensor transformations remain encapsulated within the library.

\section{Numerical validation}
\label{sec:results}
In this section, we evaluate the accuracy of the proposed numerical methodology by analyzing various benchmark two-phase viscoelastic flow problems from the literature. First, we investigate a 2D viscoelastic two-phase system in simple shear flow. Next, we examine the rising of a Newtonian bubble in a viscoelastic medium, considering both 2D planar and axisymmetric configurations. Finally, we validate the method for a 3D buoyancy-driven Newtonian bubble in a viscoelastic matrix.

\subsection{Shear-Induced Deformation of a 2D Drop}
\begin{figure}
    \centering
    \includegraphics[width=0.9\textwidth]{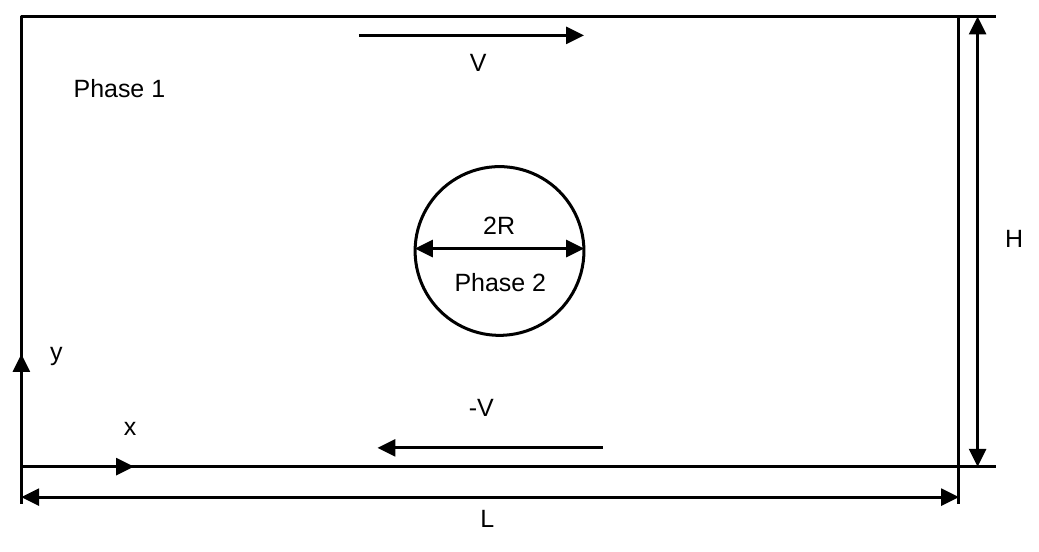}
	\caption{Sketch of the shear-driven system.}
	\label{fig:shdrivensketch}
\end{figure}

Figueiredo et al.~\cite{Figueiredo2016} investigated a 2D shear-driven droplet system in which either the dispersed or continuous phase is viscoelastic. To enable direct comparison with their results, we adopt the Oldroyd-B constitutive model. The computational domain, as sketched in Fig.~\ref{fig:shdrivensketch}, is a rectangle of size $16R \times 8R$, discretized with an equidistant mesh corresponding to 32 grid points per droplet radius $R$. A circular droplet is initially placed at the domain center. Periodic boundary conditions are imposed in the streamwise ($x$) direction, while no-slip conditions are applied at the top and bottom walls. A simple Couette shear flow ($\dot{\gamma} = V/h$) is generated by imposing opposite wall velocities, $+V$ at the top ($y = h$) and $-V$ at the bottom ($y = -h$).

The relevant non-dimensional numbers are defined as follows: the Reynolds number $Re = \rho \dot{\gamma} R^2 / (\mu^-+\eta^-)$, Capillary number $Ca = (\mu^-+\eta^-) \dot{\gamma} R / \sigma$, solvent viscosity ratio $\beta = \mu^- / (\mu^- + \eta^-)$, and Weissenberg number $Wi = \lambda^- \dot{\gamma}$. The droplet deformation is quantified using the Taylor deformation parameter, $D = (R_{\text{max}} - R_{\text{min}}) / (R_{\text{max}} + R_{\text{min}})$, where $R_{\text{max}}$ and $R_{\text{min}}$ are the maximum and minimum distances from the droplet center to the interface, respectively.

\begin{figure}[htbp]
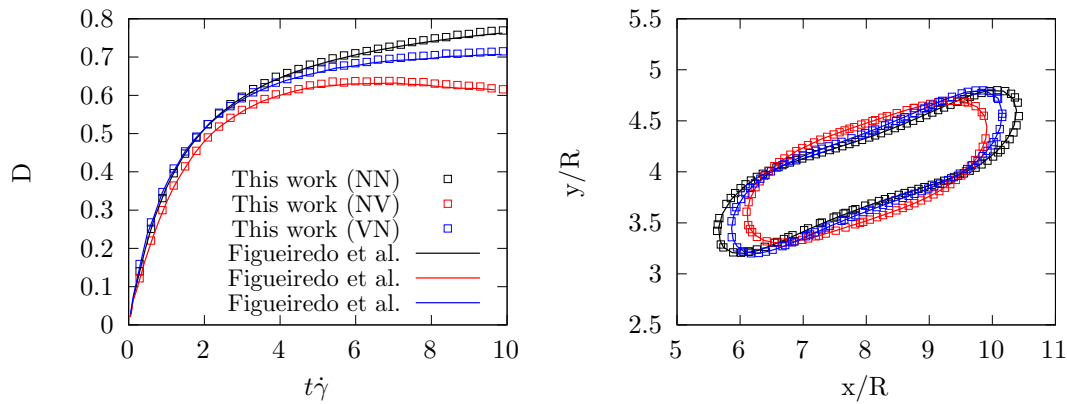

        \begin{subfigure}[b]{0.48\textwidth}
        \resizebox{\linewidth}{!}{\input{ShDrDeformation.tex}}
        \end{subfigure}%
        \begin{subfigure}[b]{0.48\textwidth}
                \resizebox{\linewidth}{!}{\input{ShDrShape.tex}}
        \end{subfigure}%
        \hfill
        \caption{2D drop deformation under simple shear flow for three configurations: all-Newtonian (NN), viscoelastic drop in a Newtonian medium (VN), and Newtonian drop in a viscoelastic medium (NV). (Left) Time evolution of the deformation parameter $D$; (Right) drop shapes at $t = 10\dot{\gamma}^{-1}$. Symbols represent the present numerical results, whereas solid lines correspond to the reference data from Figueiredo et al.~\cite{Figueiredo2016}.}
        \label{fig:2DShDr}
\end{figure}

Simulations are conducted for three cases: an Oldroyd-B droplet in a Newtonian medium (VN), a Newtonian droplet in an Oldroyd-B medium (NV), and an all-Newtonian system (NN). The parameters used are $Re = 0.3$, $Wi = 0.4$, $Ca = 0.6$, and $\beta = 0.5$. Figure~\ref{fig:2DShDr} shows the time evolution of the deformation parameter $D$ and the droplet interface shape at $t = 10\dot{\gamma}^{-1}$, alongside reference results from Figueiredo et al.~\cite{Figueiredo2016}. The comparison indicates good agreement, confirming the accuracy of the present numerical approach.

\subsection{2D Newtonian Drop Rising in a Viscoelastic Medium}
\begin{figure}
    \centering
    \includegraphics[width=0.48\textwidth]{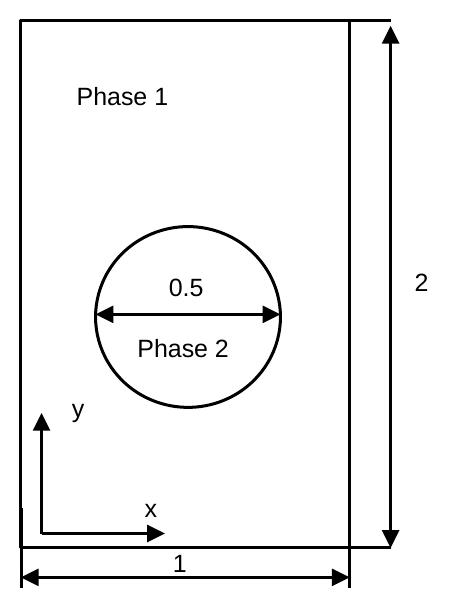}
	\caption{Sketch of the buoyancy-driven system.}
	\label{fig:bdrivensketch}
\end{figure}

Next, we validate the proposed method by simulating the rise of a Newtonian drop in an Oldroyd-B viscoelastic fluid, and comparing the results with those reported by Prieto~\cite{Prieto2015}, Vahabi and Sadeghy~\cite{Vahabi2015}, and Dritselis and Karapetsas~\cite{Dritselis2022}. The configuration (Fig.~\ref{fig:bdrivensketch}) is based on the classical Newtonian benchmark case by Hysing et al.~\cite{Hysing2009}. The computational domain has a width $L_x = 1$ and height $L_y = 2$. Initially, a circular drop of radius $R = 0.25$ is positioned at $(L_x/2, L_y/4)$. No-slip boundary conditions are applied at the top and bottom boundaries. To facilitate comparison with reference studies, different boundary conditions are imposed on the vertical sidewalls: Prieto~\cite{Prieto2015} and Hysing et al.~\cite{Hysing2009} employed free-slip conditions, whereas Dritselis and Karapetsas~\cite{Dritselis2022} and Vahabi and Sadeghy~\cite{Vahabi2015} used no-slip conditions.

The characteristic length, velocity, and time scales are defined as $L_{\text{ref}} = 2R$, $U_{\text{ref}} = \sqrt{g L_{\text{ref}}}$, and $T_{\text{ref}} = L_{\text{ref}} / U_{\text{ref}}$, respectively. The relevant non-dimensional numbers are defined as
\begin{align}
    Re &= \frac{\rho^- U_{\text{ref}} L_{\text{ref}}}{\mu^-}, \, Eo =\frac{\rho^- U_{\text{ref}}^2 L_{\text{ref}}}{\sigma}, \, Wi = \frac{\lambda^- U_{\text{ref}}}{L_{\text{ref}}}, \, \notag \\ 
    r_\mu &= \frac{\mu^+ }{\mu^- +\eta^-}, \, r_\rho = \frac{\rho^+}{\rho^-}, \, \beta = \frac{\mu^-}{\mu^- +\eta^-},
    \label{eq:numbers}
\end{align}
where $Re$, $Eo$, $Wi$, $r_\mu$, $r_\rho$, and $\beta$ denote the Reynolds number, the Eötvös number, the Weissenberg number, the viscosity ratio, the density ratio, and the solvent viscosity ratio, respectively.
 
\begin{figure}[ht!]
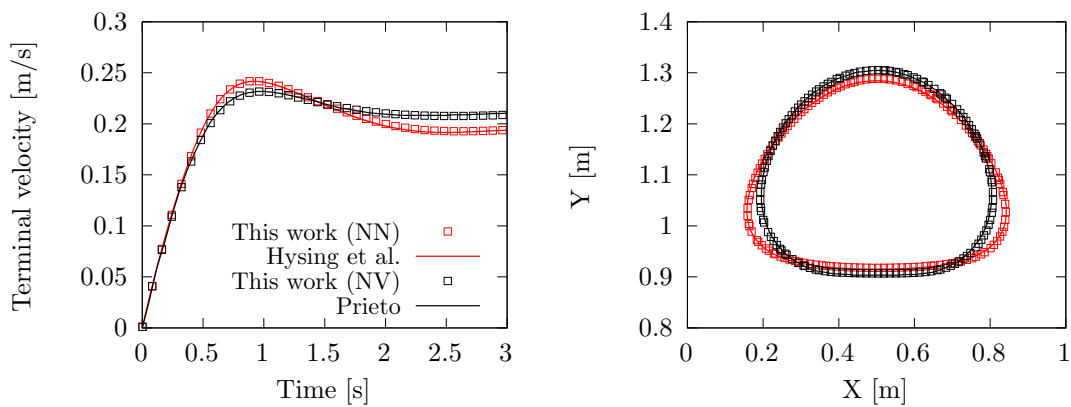

        \begin{subfigure}[b]{0.48\textwidth}
        \resizebox{\linewidth}{!}{\input{TerminalVelocity2DNNNV.tex}}
        \end{subfigure}
        \begin{subfigure}[b]{0.48\textwidth}
        \resizebox{\linewidth}{!}{\input{Shape2DNNNV.tex}}
        \end{subfigure}%
        \caption{2D rising bubble case. (Left) Evolution of the terminal velocity; (Right) steady-state bubble shapes for a Newtonian droplet in a Newtonian medium (NN) and a Newtonian droplet in a viscoelastic medium (NV).}
        \label{fig:2DRBNNNV}
\end{figure}

We first compare our results for the NN ($Wi = 0$, $\beta = 1$) and NV ($Wi = 1$, $\beta = 0.5$) cases with the reference data from Hysing et al.~\cite{Hysing2009} and Prieto~\cite{Prieto2015}, respectively. For both cases, the remaining parameters are set to $Re = 35$, $Eo = 10$, $r_\mu = 0.1$, and $r_\rho = 0.1$. A uniform Cartesian mesh consisting of $128 \times 256$ cells is employed. Figure~\ref{fig:2DRBNNNV} presents the time evolution of the terminal velocity and the steady-state bubble shapes for both NN and NV cases. It is worth noting that Prieto~\cite{Prieto2015} used the microscopic Hookean dumbbell model, in contrast to the Oldroyd-B model employed in this study. Nonetheless, the present results show good agreement with the reference data from both Hysing et al.~\cite{Hysing2009} and Prieto~\cite{Prieto2015}, validating the robustness of the numerical methodology.

Next, the numerical method is further validated for the NV case using the following parameters: $Re = 1.491$, $Eo = 35.28$, $Wi = 8.083$, $\beta = 0.07$, $r_\mu = 0.1$, and $r_\rho = 0.1$. It is noteworthy that this particular configuration has led to inconsistent results in the literature~\cite{Vahabi2015,Izbassarov2018,Dritselis2022}. Specifically, discrepancies in the droplet shape at $t = 0.13$~s have been reported.

Earlier studies by Zainali et al.~\cite{Zainali2013} and Vahabi and Sadeghy~\cite{Vahabi2015} produced different shapes at this time instant. Vahabi and Sadeghy~\cite{Vahabi2015} observed a cusped tail, a characteristic feature in buoyancy-driven NV systems with high polymer concentrations~\cite{Izbassarov2015}. In contrast, Zainali et al.~\cite{Zainali2013} reported a smoother drop shape. A later study by Izbassarov et al.~\cite{Izbassarov2018} reproduced the cusped shape reported by Vahabi and Sadeghy~\cite{Vahabi2015}. Similarly, Dritselis and Karapetsas~\cite{Dritselis2022} obtained a drop with a cusped tail; however, the droplet position was found to lie between those observed by Vahabi and Sadeghy~\cite{Vahabi2015} and Zainali et al.~\cite{Zainali2013}. Dritselis and Karapetsas reported that results at a later time ($t = 0.17$~s) were more consistent with the observations of Vahabi and Sadeghy, and attributed the earlier discrepancies to differences in numerical schemes.

\begin{figure}[ht!]
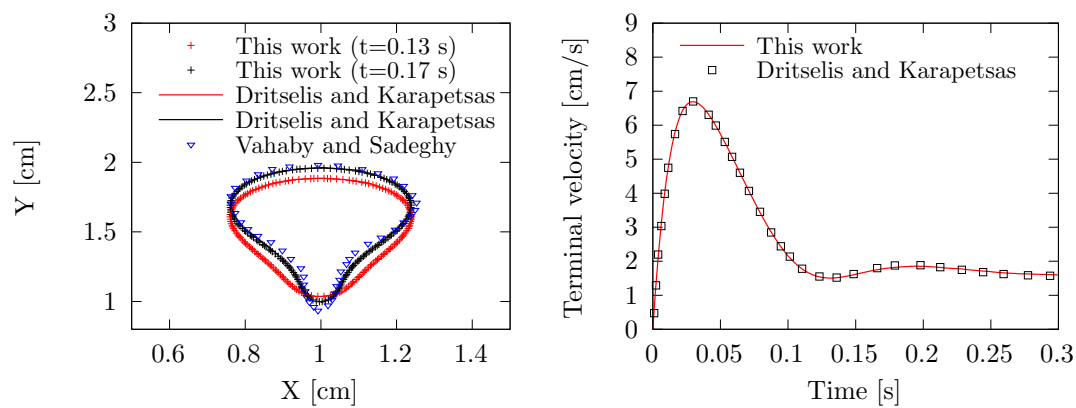

        \begin{subfigure}[b]{0.48\textwidth}
                \resizebox{\linewidth}{!}{\input{ShapeNV_T0.13_DK.tex}}
        \end{subfigure}%
        \begin{subfigure}[b]{0.48\textwidth}
                \resizebox{\linewidth}{!}{\input{TerminalVelocityNV_DK.tex}}
        \end{subfigure}%
        \caption{2D rising bubble case. (Left) Droplet shapes at $t = 0.13$ s and $t = 0.17$ s; (Right) evolution of terminal velocity for a Newtonian droplet rising in a viscoelastic medium (NV case).}
        \label{fig:2DRBNV}
\end{figure}

In the present work, the droplet shapes at $t = 0.13$~s and $t = 0.17$~s are compared with the findings of Vahabi and Sadeghy~\cite{Vahabi2015} and Dritselis and Karapetsas~\cite{Dritselis2022}, as shown in Fig.~\ref{fig:2DRBNV}. Additionally, we examine the evolution of the terminal velocity for the same configuration. Overall, the present results show good agreement with those reported by Dritselis and Karapetsas, further demonstrating the accuracy and robustness of the proposed numerical methodology.

\subsection{Axisymmetric Rising Bubble in a Viscoelastic Matrix}
To further validate our numerical methodology, we analyze buoyancy-driven dynamics in an axisymmetric configuration. Esposito et al.~\cite{Esposito2023} and Ohta et al.~\cite{Ohta2015} investigated the motion of a viscous drop in a viscoelastic medium using numerical and experimental approaches, respectively. To directly compare the results of Esposito et al., we adopted the exponential PTT constitutive model. The rheological parameters for the 1.6\% HASE polymeric solution, based on their data, are listed in Table~\ref{tab:tab2}.

\begin{figure}[ht!]
        \centering
        \includegraphics[width=\linewidth]{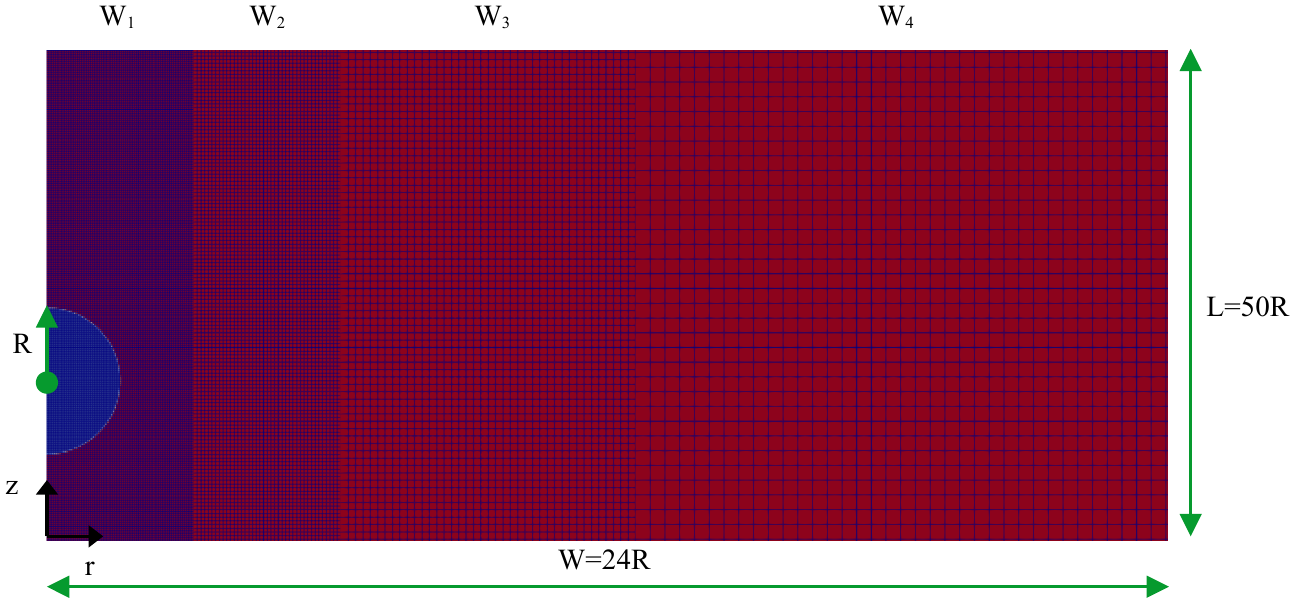}
        \caption{Axisymmetric rising bubble in a viscoelastic medium (NV case). Computational domain and mesh (not to scale), where red and blue indicate the continuous viscoelastic and dispersed phases, respectively.}
        \label{fig:AxiRBNVmesh}
\end{figure}

The computational domain and corresponding mesh are schematically illustrated in Fig.~\ref{fig:AxiRBNVmesh}. A multilayer mesh generated using the \texttt{blockMesh} utility. The mesh parameters are summarized in Table~\ref{tab:tab1}. A Newtonian drop of radius $R$ is initially positioned at $(0,\,5R)$ within a cylindrical domain of height $L = 50R$ and radius $W = 24R$. Axisymmetry is assumed, with the left boundary representing the centerline. A no-slip condition is applied to the velocity, and a zero-gradient condition is applied to the pressure, on the right and bottom boundaries. On the top boundary, atmospheric pressure is prescribed, along with a zero-gradient condition for velocity. For the viscoelastic stress tensor, Neumann boundary conditions are enforced on the lateral and top boundaries of the domain.

The relevant non-dimensional numbers are: the Weissenberg number $Wi=\lambda^-  \Delta \rho gR/(\mu^-+\eta^-)$, the Archimedes number $Ar=\rho^- \Delta \rho gR^3/(\mu^-+\eta^-)^2$, the Bond number $Bo=\Delta \rho gR^2/\sigma$,  the viscosity ratio $r_\mu=\mu^+/(\mu^-+\eta^-)$, the density ratio $r_\rho=\rho^+/\rho^-$, and the solvent viscosity ratio $\beta = \mu^-/(\mu^- +\eta^-)$.

\begin{figure}[ht!]
\centering
\input{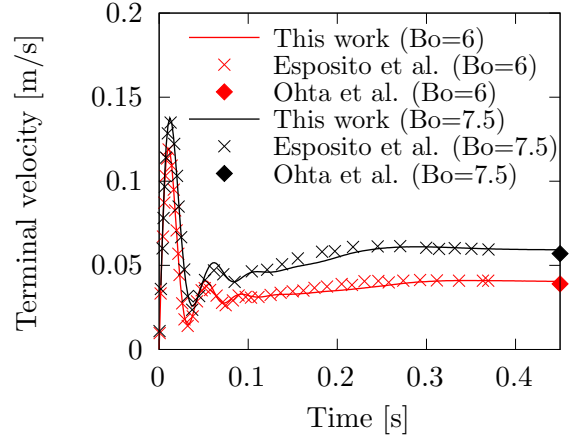}
\caption{Axisymmetric rising bubble in a viscoelastic medium (NV case). The evolution of terminal rising bubble in a viscoelastic medium (NV case). The evolution of terminal velocity for $Bo = 6$ and $Bo = 7.5$.}
        \label{fig:AxiRBNVvel}
\end{figure}

\begin{figure}[ht!]
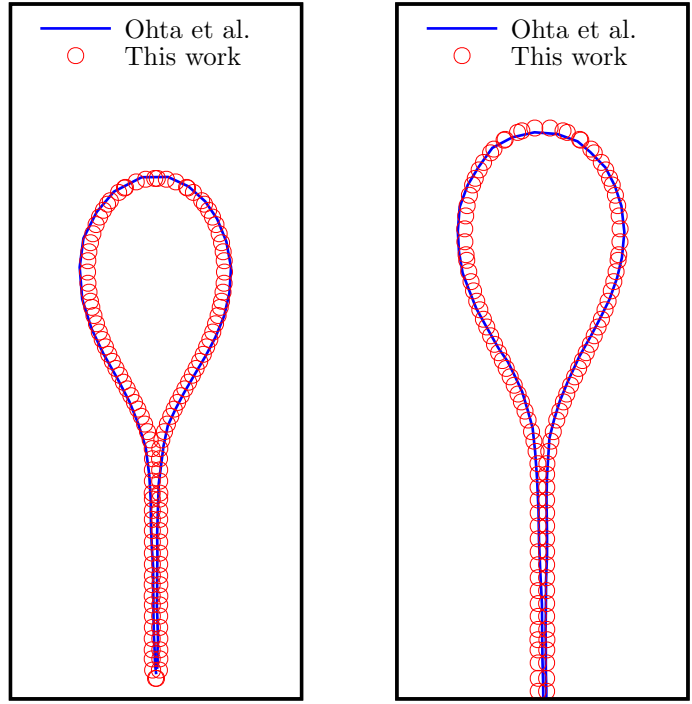

\centering
\input{ShapeBo5p7.tex}
\input{ShapeBo7p4.tex}
\caption{Axisymmetric rising bubble in a viscoelastic medium (NV case). Comparison of the present results against reference data from Esposito et al.~\cite{Esposito2023} and Ohta et al.~\cite{Ohta2015} for $Bo = 6$ and $Bo = 7.5$. The steady-state bubble shapes at $t = 0.71$ s for $Bo = 6$ and $Bo = 7.5$, respectively.}
\label{fig:AxiRBNVshape}
\end{figure}

We compare the present results at $Bo = 6$ and $7.5$ with the reference data reported by Ohta et al.~\cite{Ohta2015} and Esposito et al.~\cite{Esposito2023}, as shown in Figs.~\ref{fig:AxiRBNVvel} and \ref{fig:AxiRBNVshape}. The overall evolution of the terminal velocity shows good agreement with the numerical results of Esposito et al. At steady state, both the velocity and bubble shape converge to the experimental data reported by Ohta et al. Notably, the present results successfully reproduce the elongated and cusped tail observed experimentally by Ohta et al.

\begin{table}[t!]
    \centering
    \begin{tabular}{|c|c|c|c|c|c|c|}
    \hline
         w($\%$) & $\rho_2(kg/m^3)$ & $\mu_{s_2}$ & $\mu_p (Pas)$ & $\lambda (s)$ & $\epsilon$ & $\sigma (N/m)$ \\ 
         \hline 
         1.6 & 956 & 0.01 & 2.73 &  0.06 & 0.065 & 0.0442\\
         \hline
    \end{tabular}
    \caption{Rheological parameters of a hydrophobically modified alkali-soluble emulsion polymer (HASE)~\cite{Ohta2015} are based on the fit by Esposito et al.~\cite{Esposito2023}.}
    \label{tab:tab2}
\end{table}

\begin{table}[t!]
    \centering
    \begin{tabular}{|c|c|c|}
      \hline 
         Quantity & Length & Number of cells \\ 
         \hline
         \hline
         $W_1$ & $2R$ & 80 \\
         $W_2$ & $2R$ & 40  \\
         $W_3$ & $4R$ & 40 \\
         $W_4$ & $16R$ & 80 \\
         \hline
    \end{tabular}
    \caption{A summary of the numerical mesh.}
    \label{tab:tab1}
\end{table}

\subsection{Three-dimensional bubble rising in a viscoelastic fluid}
Next, two three-dimensional validation cases of a Newtonian bubble rising in an Oldroyd-B viscoelastic fluid were simulated. The obtained results are compared with the numerical data reported by Yuan et al.~\cite{Yuan2020} and Naseer et al.~\cite{Naseer2023} to assess the accuracy of the proposed numerical methodology.

The characteristic length, velocity, and time scales are defined as \( L_{\text{ref}} = R \), \( U_{\text{ref}} = \sqrt{gR} \), and \( T_{\text{ref}} = \sqrt{R/g} \), respectively. The associated non-dimensional numbers are provided in \cref{eq:numbers}. The computational setup is based on the configurations reported by Yuan et al.~\cite{Yuan2020} and Naseer et al.~\cite{Naseer2023}, with a domain size of \( 8R \times 40R \times 8R \) in the \(x\), \(y\), and \(z\) directions, where \(R\) denotes the initial bubble radius. The base Cartesian mesh contains \(80\times400\times80=2.56\) million cells. Dynamic AMR is applied to cells intersected by the interface every ten time steps, using two buffer layers and at most two refinement levels, with a global limit of six million cells. The bubble is centered in both lateral directions and initially positioned at \(y=3R\), corresponding to \((4R,\,3R,\,4R)\) in the archived case coordinates.

Zero-gradient boundary conditions are applied to all boundaries for both the polymeric stress and the volumetric phase fraction. For pressure, atmospheric conditions are prescribed at the top boundary, while a zero pressure-gradient boundary condition is enforced on the remaining sides. In terms of velocity, a zero-gradient condition is applied at the top, and no-slip conditions are imposed on the other boundaries. The gravitational force acts in the negative \( y \)-direction.

\begin{figure}[ht!]
        \begin{subfigure}[b]{0.6\textwidth}
                \input{Comb3D.tex}
        \end{subfigure}%
         \begin{subfigure}[b]{0.35\textwidth}                
         \includegraphics[width=\linewidth]{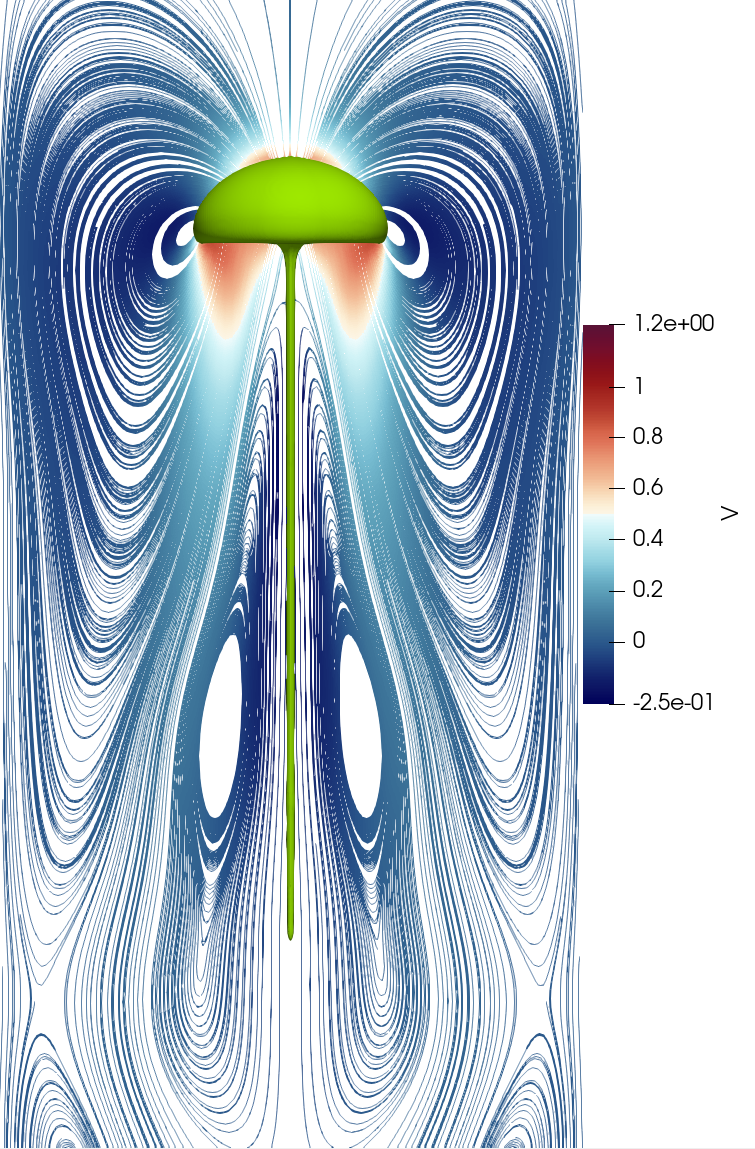}
        \end{subfigure}
\caption{
3D rising bubble in a viscoelastic medium. The left panel illustrates the temporal evolution of the terminal velocity for Case A and Case B. Present results (lines) are compared with reference data~\cite{Yuan2020, Naseer2023} (symbols). Red and blue colors represent Case A and Case B, respectively. The inset compares the steady-state bubble shape for Case A with the results of Yuan et al.~\cite{Yuan2020}. The right panel depicts the bubble shape along with the corresponding flow streamlines in a mid-plane section, with the streamlines color-coded by the local velocity magnitude.}
\label{fig:3DRBNV}
\end{figure}

Figure~\ref{fig:3DRBNV} presents a comparison between our simulation results and the reference data reported by Yuan et al.~\cite{Yuan2020} and Naseer et al.~\cite{Naseer2023}. The evolution of terminal velocity for Case A and Case B (as defined in Table~\ref{tab:tab3}) is shown on the left, demonstrating good agreement with the reference studies. The inset provides a comparison of the steady-state bubble shape for Case A against the results of Yuan et al. On the right side of the figure, the bubble shape for Case B is depicted along with streamlines colored by the local velocity magnitude. As observed in previous studies~\cite{Yuan2020,Naseer2023}, a distinct negative wake forms behind the rising bubble.

\begin{table}[t!]
    \centering
    \begin{tabular}{|c|c|c|c|c|c|c|}
    \hline
         Case & $Re$ & $Eo$ & $Wi$ & $\beta$ & $r_\mu$ & $r_\rho$ \\ 
         \hline 
         A   & 3.91 & 100 & 1 &  0.8 & 0.01 & 0.001\\
         B   & 3.91 & 100 & 10 &  0.2 & 0.01 & 0.001\\
         \hline
    \end{tabular}
    \caption{Flow parameters for two viscoelastic configurations.}
    \label{tab:tab3}
\end{table}

\section{Conclusion}
\label{sec:conclusion}


DeboRheo v2.0 extends the unstructured geometrical VOF framework of DeboRheo
to more demanding viscoelastic two-phase flows involving large property
contrasts, substantial interfacial deformation, and thin liquid structures.
The principal numerical modification concerns the treatment of solvent
viscosity at fluid interfaces. Whereas the previous formulation evaluated the
face viscosity from the geometrically reconstructed face area fraction,
DeboRheo v2.0 constructs the face mixture viscosity by separately interpolating
the cell-centered phase fraction and phase viscosities. This treatment
introduces a mesh-dependent viscous transition over the finite-volume
interpolation stencil while retaining the sharp geometrically reconstructed
VOF interface. The revised formulation substantially improves numerical
robustness when large viscosity contrasts coincide with thin viscoelastic
lamellae or filaments.

The updated solver retains geometrical interface reconstruction and phase
advection that is consistent with both mass and momentum. It also supports consistent
pressure--velocity coupling through SIMPLEC, which can be activated using the
OpenFOAM PIMPLE controls. The framework combines these numerical components
with a modular treatment of viscoelastic constitutive equations and
stabilization methods based on transformed conformation tensors.

The methodology was assessed using two-dimensional, axisymmetric, and
three-dimensional viscoelastic two-phase benchmarks governed by the Oldroyd-B
and nonlinear exponential Phan--Thien--Tanner models. The results agree well
with published numerical and experimental reference data across the
investigated configurations. The validation covers a demanding range of
free-surface dynamics, including substantial interfacial deformation,
elongated tails, and the formation of thin lamellae and filaments. These
results demonstrate the improved robustness and versatility of DeboRheo v2.0.

DeboRheo v2.0 provides an open, modular, and extensible framework for
simulating interface-resolved viscoelastic two-phase flows on unstructured
meshes. An important remaining limitation concerns the transport of the
viscoelastic constitutive state. In the present one-field formulation, the
conformation tensor, or its transformed representation, is transported
throughout the computational domain using conventional finite-volume
advection. Its transport is therefore not geometrically consistent with the
phase-specific VOF flux, allowing numerical diffusion of the constitutive state
across the interface. Future work will focus on developing a geometrically
consistent co-advection scheme for discontinuous tensor-valued constitutive
variables, systematically characterizing the mesh dependence and accuracy of
the interfacial viscosity treatment, and extending the framework to
increasingly inertia-dominated flows and topologically complex viscoelastic
free-surface configurations.

\section*{Acknowledgements}
D.I. and R.S. acknowledge financial support from the Research Council of Finland grant 354620. D.B. and T.M. acknowledge funding by the German Research Foundation (DFG)--Project ID 265191195--SFB 1194. M.M. acknowledges financial support from the Scientific and Technical Research Council of Türkiye (TUBITAK) [Grant Number 119M513]. The authors wish to acknowledge CSC – IT Center for Science, Finland, for computational resources.
 
\section*{Data and software availability}
The source code, pinned third-party dependencies, validation cases, reference
data, and step-by-step reproduction instructions are publicly available in the
immutable \href{https://gitlab.com/deborheo/deborheorelease/-/tree/v2.0.5}%
{DeboRheo v2.0.5 release}. All results reported in this work were obtained using
the released DeboRheo v2.0 code with OpenFOAM v2512. The release provides the
complete case inputs, all relevant numerical and experimental reference data
used for validation, publication-quality comparison figures, and the workflows
required to reproduce and assess the reported results.








\bibliographystyle{elsarticle-num}
\bibliography{bibliography}

\end{document}